\documentstyle[12pt]{article}
\begin{document}
\pagestyle{empty}
{}\hspace*{15em}\[
\hspace*{15em}\begin{array}{l}
\hspace*{15em}\mbox{MPI-Ph/92-118}\\
\hspace*{15em}\mbox{December 1992}
\hspace*{15em}\end{array}
\hspace*{15em}\]
\vspace*{15pt}
\begin{center}
{\sc {\bf BOTTOM--UP APPROACH TO UNIFIED}}\\
{\sc {\bf SUPERGRAVITY MODELS}}
\footnote{Supported in part by the Polish Committee for
Scientific Research}
\end{center}
\vspace*{20pt}
\begin{center}
Marek Olechowski\\
Insitute of Theoretical Physics\\
Warsaw University\\
ul. Ho\.{z}a 69, 00-681 Warsaw, Poland
\end{center}
\vspace*{10pt}
\begin{center}
Stefan Pokorski~\footnote{On leave of absence from the Institute
of Theoretical Physics, Warsaw University}\\
{\it Max--Planck--Institut f\"{u}r Physik}\\
{\it -- Werner--Heisenberg--Institut --}\\
{\it P.O.Box 40 12 12, M\"{u}nich, Fed. Rep. Germany}
\end{center}
\vspace*{30pt}
\begin{center}
ABSTRACT
\end{center}
\vspace*{10pt}

\indent A new approach is proposed to phenomenological study of a generic
unified supergravity model, which reduces to the minimal supersymmetric
standard model. The model is effectively parametrized in terms of five
low energy observables. In consequence, it is easy to
investigate systematically the parameter space of the model,
allowed by the requirement of
radiative electroweak symmetry breaking and by the present
experimental limits.

Radiative corrections due to large Yukawa couplings and
particle--sparticle mass splitting are included into the
analysis and found to have important effects, in particular on the degree of
fine tuning in the model.

In this framework there are presented the predictions of the
model for various low energy physical observables and their
dependence on the values of the top quark mass and $\tan\beta$
is discussed. Results are also given for the large $\tan\beta$ scenario,
$\tan\beta\approx m_t/m_b$.

Our approach can be easily extended to non--minimal supergravity models,
which do not assume the universality of the soft breaking parameters at the
unification scale $M_X$. Such an extension will be particularly useful
once the masses of some sparticles are known, allowing for a model independent
study of the parameter space at $M_X$.
\newpage
\pagestyle{plain}
\begin{flushleft}
{\sc {\bf 1.~~Introduction}}
\end{flushleft}
\renewcommand{\theequation}{1.\arabic{equation}}
\setcounter{equation}{0}

\indent The "supergravity inspired" minimal supersymmetric
standard model (MSSM) is an interesting laboratory for studying
various phenomenological aspects of supersymmetry. In this model
the soft supersymmetry breaking parameters of the low energy
lagrangian are given, via the renormalization group (RG)
evolution, in terms of the unification scale parameters. These
are: the universal gaugino and scalar  masses $M_o$ and $m_o$, respectively,
and
the trilinear scalar coupling $A_o$. Apart from the gauge and Yukawa
couplings there are in the model two more free parameters: the Higgs
superfield mixing mass $\mu_o$ in the superpotential and the
corresponding soft breaking parameter $B_o$. The actual
mechanism of the soft supersymmetry breaking (and
the origin of the $\mu$--term in the superpotential) are not yet
understood theoretically. Although there are various theoretical
suggestions with regard to the values of those parameters at
the unification scale, they should be considered at present as
free parameters of the model.

One of the most attractive features of the supergravity models
is the possibility of radiatively induced gauge symmetry breaking$^{[1]}$,
which becomes manifest at the tree level after the RG evolution of
the parameters of the lagrangian from the unification scale
$M_X$ to the scale $M_Z$.
This way the hierarchy $M_Z/M_X$ becomes related to the presence in
nature of large Yukawa couplings. The minimal model supplemented
by the requirement of radiative gauge symmetry breaking is a
very predictive framework: with only 5 free mass parameters
$\mu_o, m_o, M_o, A_o$ and $B_o$ (and the yet unknown top quark
Yukawa coupling $Y^o_t$), it interconnects 1)~the $SU(2)\times U(1)$
gauge symmetry breaking with
$v$=174 GeV, 2)~the very constrained Higgs sector of the MSSM, 3)~the
sparticle mass spectrum and 4) the relic density of the lightest
supersymmetric particle (LSP). In
addition, it has recently been strongly reemphasized that
the predictivity of the model is greatly enhanced by the requirement
of no fine tuning of the parameters$^{[2]}$. Indeed,
the requirement that the $M_{W,Z}$ be within the measured value
without fine tuning of the unification scale parameters strongly
limits the magnitude of the supersymmetry breaking scale and, in
consequence, of the sparticle masses.

In spite of the large number of papers devoted to the model, its analysis
is not yet fully satisfactory (for a complete list of earlier
references we refer the reader to the ref. [3]). Its full parameter space
has not been systematically explored yet. Secondly, it has
recently been understood that radiative corrections originating
from the combined effect of large Yukawa couplings and
particle--sparticle mass splitting have important effects on the
mechanism of  radiatively induced $SU(2)\times U(1)$ symmetry
breaking$^{[4]}$ and on the Higgs sector$^{[5]}$. In consequence, one may
expect important modifications in the predictions of the model.

There are two main points in our strategy in this paper.
Firstly, we set up the formalism to explore the full parameter
space of the model. Our formalism is based on the bottom to top
approach whose essence is as follows: we choose a set of values for 5 low
energy physical parameters, and then look for the set of parameters
at the unification scale $M_X$ which induces radiative gauge symmetry
breaking and gives the chosen low energy values. Thus, we reverse
the standard procedure which is to find the low energy
parameters corresponding to a given set of the boundary
conditions at the unification scale. This way we can systematically
study the phenomenology of the "supergravity
inspired" MSSM compatible with the idea of
radiatively induced symmetry breaking, with no additional
assumptions (apart from their universality) about the values (and their
correlation)
of the parameters at
$M_X$ (for which we have little theoretical insight). In this approach it is
straightforward to implement all the experimental constraints on the low energy
parameters and to discuss systematically the correlations between
physical observables, which are predicted by the model.
{\it Effectively, the model is now parametrized by five low energy
physical parameters.} We would also like to stress that our
approach is readily extendable to non--minimal supergravity
models, which do not assume the universality of the soft
supersymmetry breaking terms at the scale $M_X$, provided more
than five low energy physical parameters is used as input. Once
some sparticles are discovered and their masses known, such an
extension will allow for a model independent study of the
parameter space at the unification scale.

Secondly, we consistently include radiative corrections generated by large
Yukawa couplings and particle--sparticle mass splitting. To
this end, we use the  RG equations for all  parameters
of the potential with heavy sparticles decoupled at their
thresholds. In the context of  radiative breaking of the
$SU(2)\times U(1)$ symmetry this technique has already been
advocated in ref.[6]. It has also been used to study similar corrections
to the Higgs sector in the MSSM and found to be quite accurate$^{[7]}$.
In this paper we use the RG equations of ref.[6] with  squarks and
gluinos decoupled at $M_{SUSY}$ (the precise definition of this
scale will be discussed in Section 2). Radiative corrections generated
by the other sparticle--particle mass splitting are small
and can be neglected.

The question about the role of the large radiative corrections
in the phenomenological analysis of the model
has already been given some attention in the literature (see
ref. [3] and references therein). However, to our knowledge, no
fully consistent treatment of such corrections at the 1--loop level,
simultaneously in all  four sectors mentioned above, exists yet. Their full
impact on the phenomenology of the model has become easy to
explore only within our bottom to top approach.

The paper is organized as follows: In the next Section we
present the details of our approach. Section 3 is devoted to a
qualitative discussion of those predictions of the model which
follow from the requirement of radiative gauge symmetry
breaking. In Section 4 we discuss the problem of fine tuning of
the parameters and emphasize the role of radiative corrections
in sizably diminishing the degree of fine tuning. In Section 5
we present our results, after imposing the presently available
experimental constraints. Finally, Section 6 contains a very
brief discussion of the large $\tan\beta$ scenario and a summary
of this work is given in Section 7.\\
\vspace*{5pt}
\begin{flushleft}
{\sc {\bf 2.~~The Bottom--Up Approach}}
\end{flushleft}
\renewcommand{\theequation}{2.\arabic{equation}}
\setcounter{equation}{0}
The details of our procedure are as follows. We begin
with choosing, in addition to the "experimentally" known
quantities
\begin{eqnarray}
&&\alpha_{EM}(M_Z)=1/127.9~,\hspace{5mm}\sin^2\theta_W
(M_Z)=0.2324~,\nonumber\\
&& m_{\tau}(M_Z)=1.746~\mbox{GeV},\hspace{5mm} M_Z=91.19~\mbox{GeV}~,
\end{eqnarray}
a set of values for the following low energy unknown parameters:
\begin{equation}
M_t,~~M_A,~~\tan\beta,~~m^2_Q,~~m^2_U
\end{equation}
where $M_A$ is the Higgs pseudoscalar mass and $m^2_Q$ and
$m^2_U$ are the soft (low energy) mass parameters of the squarks
of the third generation. The physical squark masses are
eigenvalues of the mass matrix. For the bottom to top approach
this is the most economical choice of the low energy input
parameters: $M_Z,~~\tan\beta$ and $M_A$ are directly related to
the parameters of the tree level potential and the squark mass
parameters determine the magnitude of the dominant radiative
corrections to the scalar quartic couplings. So they must be
fixed anyway at the begining of the analysis. To proceed further
we have to discuss in some detail the unification of gauge and
Yukawa couplings.
\begin{flushleft}
{\bf 2.1.~~Unification of gauge and Yukawa couplings}
\end{flushleft}

The analysis of the unification of couplings in the MSSM
have been addressed in a number of papers$^{[8]}$, with the most
recent and thorough discussion given in ref. [9,10]. At the 2--loop
level, the RG equations for the gauge and Yukawa couplings are a
coupled set of equations. However, in the calculation of the
gauge couplings, it is a very good approximation to set the top
quark Yukawa coupling to be constant$^{[9]}$ (for
$\tan\beta\approx m_t/m_b$, the b-quark Yukawa coupling is
as large as the t-quark coupling and the same procedure must be
applied to both of them). Then the
effect of the Yukawa couplings on the running of $\alpha_i$ is
realized as a negative correction to the 1--loop coefficients
$b_i$'s (which is, anyway, small; see below). Effectively, for
some chosen value of $Y_t$, we can
first discuss the gauge coupling unification. Starting from the
experimental values for $\alpha_{EM}(M_Z)$ and
$\sin^2\theta_W(M_Z)$, we use the unification conditions to
predict the values of the running strong coupling
constant $\alpha_3(Q)$ and the unification scale $M_X$ (for the
purpose of this paper we neglect the experimental uncertainity
in the determination of $\sin^2\theta_W$). After the inclusion of
the 1--loop threshold corrections coming from the supersymmetric
particles, we obtain for each gauge coupling the relation
\begin{equation}
\frac{1}{\alpha_i(Q)}=\frac{1}{\alpha_i(M_X)}+\frac{b_i^{~MSSM}}
{2\pi}~\ln\left(\frac{M_X}{Q}\right)+\gamma_i+\frac{1}{\alpha_i^{~th}}
+\Delta_i(Q)
\end{equation}
where $\gamma_i$ contains the 2--loop contributions to the beta
functions, and
\begin{equation}
\frac{1}{\alpha_i^{~th}}=\sum_{\eta,~M_{\eta}>Q}~\frac{b_i^{\eta}}
{2\pi}~\ln\left(\frac{M_{\eta}}{Q}\right)
\end{equation}
is the 1--loop threshold correction to 1/$\alpha_i(Q)$. The
summation is over all sparticle and heavy Higgs doublet states
with masses above $Q;~b_i^{\eta}$ is the contribution of each of
those particles to the 1--loop beta function coefficient of the gauge
coupling $\alpha_i$ (when solving the RG equations at the
2--loop level, it is sufficient to consider threshold
corrections at the 1--loop level). The $\Delta_i$ includes the
threshold correction related to the top quark mass and the
effect of the $Y_t$ on the running of $\alpha_i$. The importance of
various terms in eq. (2.3) can be illustrated by considering
their contribution to the value of $\alpha_3(M_Z)$, obtained
from the unification conditions. It is well known that the
2--loop corrections $\gamma_i$ give $\sim$10\% increase in the
value of $\alpha_3(M_Z)$ as compared to the 1--loop result and
must be included into the analysis. The 2--loop top quark Yukawa
coupling corrections to $\alpha_3(M_Z)$, included in
$\Delta_i$'s, give $-0.001,~-0.0015,~-0.002,~-0.003$ for $Y^2_t(M_X)/4\pi
\simeq 0.1,~0.2,~0.4,~1.0$, respectively$^{[10]}$ and the top quark
threshold correction is even smaller, (and of opposite sign)
or at most similar, for $M_t<200~\mbox{GeV}^{[9]}$. Altogether, for the range
of the top quark masses and Yukawa couplings considered in this paper, those
corrections do not exceed (1--2)\% and we neglect them in our analysis.

The contribution of the sparticle thresholds to
$1/\alpha_3(M_Z)$ can be effectively parametrized in terms of
only one parameter $T_{SUSY}$ as follows:
\begin{equation}
\Delta^{th}\left(\frac{1}{\alpha_3(M_Z)}\right)=\frac{1}{2\pi}~\ln~
\frac{T_{SUSY}}{M_Z}
\end{equation}
where the scale $T_{SUSY}$ has the obvious interpretation in the
(unrealistic) case of the fully degenerate sparticle spectrum:
it is just the common mass of all the sparticles. Notice that,
in this case, a non--zero threshold corrections to
$1/\alpha_i(M_Z)$ are present only if $T_{SUSY}>M_Z$ and, in
consequence, the contribution (2.5) implies a decrease in the
value of $\alpha_3(M_Z)$.

In the realistic case, in the presence of sparticle mass
splitting, and in the approximation in which one neglects the
effects of the electroweak symmetry breaking on the sparticle
masses, i.e. assuming no mixing, the scale $T_{SUSY}$ reads$^{[10]}$:
\begin{equation}
T_{SUSY}=m_{\tilde{H}}\left(\frac{m_{\tilde{W}}}{m_{\tilde{g}}}\right)^
{28/19}\left(\frac{m_{\tilde{l}}}{m_{\tilde{q}}}\right)^{3/19}
\left(\frac{m_H}{m_{\tilde{H}}}\right)^{3/19}\left(\frac{m_{\tilde{W}}}
{m_{\tilde{H}}}\right)^{4/19}
\end{equation}
where $m_{\tilde{q}},~m_{\tilde{g}},~m_{\tilde{l}},~m_{\tilde{W}},~
m_{\tilde{H}}$ and $m_H$ are the characteristic masses of the
squarks, gluinos, sleptons, electroweak gauginos, Higgsinos and
the heavy Higgs doublet, respectively. Here we have assumed that all the
sparticle masses are larger than $M_Z$ (in the approximation of
no mixing); if any sparticle mass is smaller then $M_Z$, it
should be replaced by $M_Z$ when applying eq. (2.6). It is
obvious that $T_{SUSY}$ depends strongly only on the first two
factors and in the model under investigation, with gaugino
masses coming from a common supersymmetry breaking scale $M_o$
at $M_X$, $T_{SUSY}$ is well approximated by$^{[10]}$:
\begin{equation}
T_{SUSY}\approx m_{\tilde{H}}\left(\frac{\alpha_2(M_Z)}{\alpha_3(M_Z)}
\right)^{3/2}\approx\mu~\left(\frac{\alpha_2(M_Z)}{\alpha_3(M_Z)}\right)^{3/2}
\end{equation}
Clearly, unless the Higgs mixing parameter in the superpotential
$\mu>600~\mbox{GeV}$, the effective scale $T_{SUSY}<M_Z$ and the
sparticle threshold corrections tend to increase the value of
$\alpha_3$. As it will be discussed later on, large values of
$\mu$ are less natural and the most interesting range is
$100~\mbox{GeV}<\mu<600~\mbox{GeV}$. Therefore, for our model, the typical
values are $15~\mbox{GeV}<T_{SUSY}<M_Z$. The corresponding values of
$\alpha_3(M_Z)$
obtaind from the unification condition vary from
$\alpha_3(M_Z)=0.123$ for $T_{SUSY}=M_Z$ (i.e. effectively, threshold
corrections to $\alpha_3(M_Z)$ cancell out) to
$\alpha_3(M_Z)=0.127$ for $T_{SUSY}=15~\mbox{GeV}$ (this case can be
realized in practice e.g. by decoupling the colour sparticles
in the RG equations for $\alpha_i$'s at $m_{col}\approx 3M_Z$
and the remaining sparticles at $M_Z$ (see eq, (2.6))). Experimentally,
$\alpha_3(M_Z)$ is poorly determined and, conservatively, any
value in the range $(0.12\pm 0.01)$ is still acceptable. We see
from the above discussion that the sparticle spectra predicted
by the model push us into the upper range of values for
$\alpha_3(M_Z)$, unless we take less natural solutions. Of
course, for full consistency, one should take care of the fact
that the threshold corrections correlate the value of $\alpha_3$
with the specific solution obtained for the parameters of the
model ($\mu$ in particular) which, in turn, slightly depend on the value of
$\alpha_3(M_Z)$. However, there is an inherent uncertainity in
the theoretical prediction for $\alpha_3(M_Z)$ which follows from the
unknown contributions from the thresholds at the unification
scale and from the higher dimension operators which are likely
to be as important as the dependence on $T_{SUSY}$$^{[9]}$.
Therefore, in this paper we take $T_{SUSY}=M_Z$ and use for the
gauge couplings the supersymmetric RG equations down to the
scale $M_Z$. This gives $\alpha_3(M_Z)=0.123$, close to the
central experimental value. For the unification scale we get $M_X\cong
2.4\times 10^{16}~\mbox{GeV}$.

Once the running of the gauge couplings is determined, we can
proceed with the determination of the top, bottom and $\tau$
Yukawa couplings in the whole range $(M_Z,~M_X)$. In our
procedure the $Y_{\tau}$ and $Y_t$ at $M_Z$ are fixed by the
values of the physical $\tau$ mass $M_{\tau}$ and the chosen
physical top quark mass $M_t$ and the value of $\tan\beta$. The
running mass $m_t(M_t)$ is related to the physical mass (defined
as the pole of the propagator) by the following 1--loop formula$^{[11]}$:
\begin{equation}
m_q(M_q) = \frac{M_q}{1+\frac{4\alpha_3(M_q)}{3\pi}}
\end{equation}
For the running of $m_t(M_t)$ down to $M_Z$ we use the 2--loop RG equations
of the MSSM and we obtain $Y_t(M_Z)$ from the equation
$m_t(M_Z)=Y_t(M_Z)v/\sqrt{1+\cot^2\beta}$. The running of the
couplings $Y_t,~Y_b,~Y_{\tau}$ in the range $(M_Z,~M_X)$ is
obtained by solving the 2--loop coupled RG equations with the
boundary condition $Y_b^o=Y^o_{\tau}$ at $M_X$. The
solutions to the RG equations with those mixed (low and high
energy) boundary conditions are obtained by iteration. For
consistency with the RG equations for the gauge couplings we use
the supersymmetric RG equations for the Yukawa couplings in the
whole range $(M_Z,~M_X)$. Indeed, it has recently been shown$^{[10]}$
that the same effective scale, $T_{SUSY}$, governs the
supersymmetric particle threshold corrections to the gauge and
Yukawa couplings.

The reason for taking the top and not the bottom mass as the
input parameter is as follows: as soon as the top quark is
discovered, its mass will be known with good precision whereas
the physical bottom mass is not directly accessible to
experiment and its value will remain very uncertain and model
dependent. In this paper we present the results for the following
sets of $(M_t,~\tan\beta)$ values: a) 130, 2, b) 130, 10, c)
180, 2, d) 180, 30, e) 130, 31 and f) 180, 51 (all masses
given in GeV). It is interesting to comment on the predictions
for the bottom quark mass obtained in each of those cases. They
are summarized below:
\begin{center}
\begin{tabular}{c|ccc}\hline
{}~&$m_b(M_Z)$&$m_b(M_b)$&$M_b$\\ \hline
a&4.01&5.57&6.08\\
b&4.04&5.60&6.12\\
c&3.30&4.71&5.16\\
d&3.55&5.02&5.49\\
e&3.81&5.33&5.82\\
f&3.17&4.56&5.01\\
\hline
\end{tabular}
\end{center}
\noindent The running mass $m_b(M_b)$ is obtained from $m_b(M_Z)$
by means of the $SU(3)\times U(1)$ 2--loop RG equations and the physical
mass $M_b$ is given by the eq. (2.8).

As said above, the physical bottom mass is a poorly known
quantity but, still, following the particle data book it is
expected to be in the range, say
\begin{equation}
M_b = 4.6-5.3~\mbox{GeV}
\end{equation}
We see that the conditions of the gauge coupling unification and
of the unification of the bottom and $\tau$ Yukawa couplings
result in too large a value of the $M_b$, for moderately heavy
top quark. It is so because the large value of $\alpha_3(M_Z)$
($\sim 0.12$) gives too much of a strong correction to the
running of $m_b(M_X)=m_{\tau}(M_X)$, which has to be partly
cancelled out by large, negative contribution from the top quark
Yukawa coupling. In consequence, as discussed in detail in ref.
[10], in this supersymmetric grand unification scenario, the value
of $M_b$ in the range (2.9) implies the top quark mass
remarkably close to its quasi infrared fixed point value
associated with the "traviality" bounds on the top Yukawa
coupling. A top quark mass $M_t\approx 130-140 ~\mbox{GeV}$ happens to
be close to its infrared fixed point (i.e. consistent with
(2.9)) only for values of $\tan\beta$ close to one, or very
large values of $\tan\beta$, $\tan\beta\geq
m_t(m_t)/m_b(m_t)$, above the value for which the unification
of the three Yukawa
couplings, $Y_t^o=Y_b^o=Y^o_{\tau}$, occurs. Those values of
$\tan\beta$ are the lower and upper (or even exeeding it) limits,
respectively, for
which radiative gauge symmetry breaking can be realized in the
model and, at best, at the expense of a rapidly increasing
fine--tuning of the parameters (to be discussed later on). The
situation is different for $M_t\sim (180-190)~\mbox{GeV}$, which
corresponds to its quasi infrared fixed point value for a large
range of $\tan\beta$ values.

In summary, exact unification of gauge and $\tau$ and the bottom
Yukawa couplings in the MSSM, together with the range (2.9) for
the bottom mass, imply that $M_t<180~\mbox{GeV}$ can be realized only for
$\tan\beta\approx 1$ or $\tan\beta\approx m_t(m_t)/m_b(m_t)$,
i.e. for values for which the mechanism of radiative gauge
symmetry breaking can be realized only at the expense of a large
amount of fine--tuning (or, in some cases not realized at all).
{}From this point of view, the values
$M_t>180~\mbox{GeV}$ look more natural in the model. However, one
should remember that the exact unification of couplings may be
subject to the high energy threshold and higher dimension
operator corrections, which can modify the above strong
conclusions. This is true for both the gauge coupling and the
$\tau$ and bottom Yukawa coupling unification condition.
\noindent For instance, if $\alpha_3(M_Z)$ is smaller by $\Delta=
-0.006$ (the maximum magnitude of the high energy corrections,
estimated in ref. [9]) then we get the following results:
\begin{center}
\begin{tabular}{c|ccc}\hline
{}~&$m_b(M_Z)$&$m_b(M_b)$&$M_b$\\ \hline
a&3.88&5.30&5.75\\
b&3.91&5.34&5.80\\
c&2.97&4.21&4.60\\
d&3.42&4.76&5.18\\
e&3.70&5.09&5.53\\
f&3.08&4.35&4.74\\
\hline
\end{tabular}
\end{center}
On the other hand, with $\alpha_3(M_Z)=0.123$ and relaxing the
bottom and $\tau$ Yukawa coupling unification, for $M_b=5.2~\mbox{GeV}$
one gets the following results for the ratio $r=h_b(M_X)/h_{\tau}(M_X)$:
\begin{center}
\begin{tabular}{c|c}\hline
{}~&r\\ \hline
a&0.829\\
b&0.820\\
c&1.008\\
d&0.923\\
e&0.846\\
f&1.135\\
\hline
\end{tabular}
\end{center}
The above examples illustrate the order of magnitude of the
corrections to the exact unification conditions, which are
necessary to make moderate values of the top quark mass and
$\tan\beta$ compatible with the range of values (2.9) for the
physical bottom mass.

In view of all the uncertainities discussed above, in this paper
we present the results for all cases (a -- f), with the gauge
and Yukawa couplings fixed by the exact unification condition.
Even in the presence in some of those cases of the likely inconsistency
with the experimental value in the predicted bottom quark mass, this is a
useful exercise since it allows to discuss interesting
correlations present in the model, as a function of $M_t$ and $\tan\beta$.
Moreover, modifications to $\alpha_3(M_Z)$ and/or to the relation
$Y_{\tau}(M_X)=Y_b(M_X)$, which are sufficient to bring down the value of
$M_b$ into the range (2.9), have only little impact on the rest of our results.
\begin{flushleft}
{\bf 2.2.~~Determination of the parameters in the Higgs potential}
\end{flushleft}
After those preliminaries we are now able to fix all the
parameters in the Higgs potential at the scale $M_Z$
\begin{eqnarray}
V&=&\hat{m}^2_1\mid H_1\mid^2+\hat{m}^2_2\mid H_2\mid^2-\hat{m}^2_3
(\epsilon_{ab}H_1^aH_2^b + h.c.)\nonumber\\
&&+\frac{1}{2}\lambda_1\mid H_1\mid^4+\frac{1}{2}\lambda_2\mid H_2\mid^4
+\lambda_3\mid H_1\mid^2 \mid H_2\mid^2\nonumber\\
&&+\lambda_4\mid \epsilon_{ab}H_1^a~H_2^b\mid^2
\end{eqnarray}
in terms of the values taken for the set (2.1) and (2.2). (As
usual the boundary values of the parameters at $M_{SUSY}$ read:
$\hat{m}^2_{1,2}=m^2_{1,2}+\mu^2,~~\hat{m}^2_3=B\mu$ where
$m^2_{1,2}$ are the soft Higgs mass parameters, $\mu$ is the
superhiggs mixing parameter and $B$ is the corresponding soft
breaking parameter, $\epsilon_{12}=+1$ and
$\hat{m}^2_3$ is defined to be positive).
However, before we proceed further, it is useful to recall the
conventional approach to radiative $SU(2)\times U(1)$ breaking.
It makes use of the renormalization group improved tree--level
potential evaluated at a renormalization scale $Q\sim M_Z$. The
running parameters in the potential, which are functions of the
renormalization point $Q$, can be evaluated for any $Q$ by
solving the standard supersymmetric RG equations (written in the
Landau gauge and in the (mass--independent) modified minimal
subtraction scheme, $\overline{MS}$) with some universal boundary
conditions at the scale $M_X$. For $Q\sim M_Z$, those RG
equations sum up all the large logarithmic corrections
proportional to $\ln (M_X/Q)$, which are reabsorbed in the
running parameters of the potential (2.10). Due to the
supersymmetric structure of the RG equations the tree--level
potential maintains its supersymmetric form at any $Q$.
Therefore, also at $M_Z$, the parameters satisfy the
supersymmetric relations:
\begin{equation}
\lambda_1=\lambda_2=\frac{1}{4}~(g_1^2+g^2_2),~~\lambda_3=\frac{1}{4}~
(g^2_2-g^2_1),~~\lambda_4=-\frac{1}{2}~g^2_2
\end{equation}
This procedure neglects radiative corrections generated by the
soft supersymmetry breaking (i.e. neglects logarithmic
corrections of the type $\ln (m_{stop}/Q)$), which are
proportional to large Yukawa couplings and as we know
now, can be very important and are of our concern here. One of
the well known consequences is that the minimum of the tree--level
potential is very sensitive to the choice of the renormalization
scale $Q$. The choice $Q=M_Z$ gives, in general, unreliable
results. The reason is that the loop correction $\Delta
V_1(M_Z)$ to the potential is not necessarily
small.\footnote{The correct choice of the scale $Q$ is such that
$\Delta V_1(Q)\approx 0$ but it is unknown a priori.} This has
been demonstrated explicitly$^{[4]}$: the sum $V_o(Q)+\Delta
V_1(Q)$ is indeed scale--independent, up to 2--loop corrections
which turn out to be small. For the sake of  easy reference
we give in the Appendix the 1--loop corrections to the
parameters of (2.10): $\Delta m^2_i(Q)$ and $\Delta\lambda_i(Q)$
(calculated in the symmetric phase$^{[6]}$ i.e. before
$SU(2)\times U(1)$ breaking) generated by the squark loops. This is the
dominant part of the entire 1--loop contribution to the
effective potential.

For our purpose, the inconvenience
of this approach is that one cannot work directly with the
tree--level potential (2.10) and therefore the relations between
the {\it physical} low energy parameters and the
unification scale initial conditions are complicated.
Fortunately, there exists the well--known equivalent method of
including radiative corrections in question, which is
based on the Appelquist--Carazzone decoupling theorem. This is
the tool to absorb in the running parameters of the tree--level
potential (2.10) also the corrections generated by the
particle--sparticle mass splitting. The main source of large
corrections is in this case the top--stop splitting. At this
point one should stress that sparticle thresholds contribute
differently to different quantities and the dominant corrections
to the scalar potential have different origin than the
corrections to the gauge couplings. Therefore, to a very good
approximation, the running parameters in (2.10) at the scale
$M_Z$ can be related to their initial values at the unification
scale by solving the RG equations in two steps: the supersymmetric
equations in the range $(M_{SUSY},~M_X)$ and the RG equations
with squarks and gluinos decoupled$^{[6]}$ at the common scale
$M_{SUSY}$, in the
\newpage
\noindent range $(M_Z,~M_{SUSY})$.\footnote{The mass
difference between the third and the remaining two generations
of squarks can be neglected as the latter contribute little, due
to the small Yukawa couplings. Gluino mass is, in most cases, close
to the squark masses. The other sparticles are kept in the
spectrum down to $M_Z$. The reason for using in this paper the
RG equations with only squarks and gluinos decoupled is that, on
the one side, radiative corrections to the scalar potential
from the other sparticles are
very small and, on the other side, we want to analyze the light
sparticle spectrum predicted by the model. However, as far the
scalar potential is concerned, very similar results are obtained
by using below $M_{SUSY}$ the RG equations of the standard model
with two Higgs doublets.} By decoupling squarks at $M_{SUSY}$
chosen to be equal to the {\it low energy} squark soft mass
parameters one reabsorbes the logarithmic parts of the
corrections (A.7) to (A.16) (in the approximation
$m^2_Q=m^2_U=m^2_D$) and, of course, also higher order leading
logarithmic corrections, into the running parameters of the
potential (2.10). As long as we work in the region ($M_Z,~M_{SUSY}$)
with the 1--loop RG equations, this is a consistent
approximation: all non--logarithmic corrections in eqs.
(A.7--A.16) are formally the effects of the same order as the
neglected 2--loop effects in the RG equations. However, earlier
results$^{[4]}$ strongly indicate that the former are much more
important than the latter. In particular, as discussed earlier,
they are important for the potential to be renormalization scale
independent (in our case, this means, independent of the choice
of $M_{SUSY}$ in a reasonable range around the soft squark masses).
Also, it is known that the 2--loop effects in the
RG equations for the evolution of the quartic couplings are very
small$^{[12]}$. Therefore, in this paper we include also the
non--logarithmic part of the 1--loop corrections (A.7--A.16) by
properly choosing the matching conditions for the parameters
$m^2_i$ and $\lambda_i$ at $M_{SUSY}$. More specifically, we fix
the scale $M_{SUSY}$ so that the contribution (A.8a) to $\Delta
m^2_2$ vanishes. All 1--loop corrections to the quartic
couplings are then included by replacing at $M_{SUSY}$ the
continuous matching conditions given by the relations (2.11) by
the conditions $\lambda_i\rightarrow\lambda_i+\Delta\lambda_i$
with $\Delta\lambda_i$'s given by the eqs. (A.13--A.16)
(everywhere in this paper we use the approximation $m^2_{U_{1,2}}=
m^2_{Q_{1,2}}=m^2_{D_{1,2}}=m^2_{D_3}=2m^2_{Q_3}-m^2_{U_3}$ which
follows from the supersymmetric RG equations after neglecting
terms proportional to the electroweak gauge couplings).
Similarly for $m^2_i,~i=1,2,3$~, we take
discontinuous matching conditions at $M_{SUSY}~:~m^2_i+\Delta
m^2_i$ where $\Delta \hat{m}^2_3$ is given by the eq. (A.9), $\Delta m^2_2$
-- by the (A.8b) and $\Delta m^2_1$ is given by the expression
(A.7b), always with $\Lambda=M_{SUSY}$.\footnote{We neglect the part (A.7a)
of $\Delta m^2_1$ which is very small for our choice of $M_{SUSY}$. One should
also remember that always $m^2_1\gg m^2_2$ and the correction
$\Delta m_1^2$ is relatively much less important than $\Delta m^2_2$.}
The technical way of implementing the
discontinuous matching conditions for $m^2_i$'s is discussed
after eq. (2.17).

After the decoupling of the sparticles the largest 1--loop
corrections to the tree level potential obtained after the
evolution are given by the top quark loops. The top quark should
therefore be decoupled at $\Lambda_{top}$ chosen to minimize
those 1--loop corrections to the vevs $v_i$:
\[
\left.\frac{\partial\Delta V_1}{\partial v_i}\right |_{v_i=v^o_i}=0
\]
where
\[
\Delta V_1=\frac{3}{(4\pi)^2}Y^2_tv^4_2\left( -\frac{3}{2}~+~\ln~\frac
{Y_t^2v_2^4}{\Lambda^2_{top}}\right)
\]
We get then $\Lambda_{top}\sim M_t/\sqrt{e}$ and for the
interesting range of the top quark masses we can use the
approximation $\Lambda_{top}\approx M_Z$. Thus, by the RG
evolution in the range $(M_Z,~M_{SUSY})$ one reabsorbes in the
tree level parameters the remaining large corrections to the potential.
\begin{flushleft}
{\bf 2.3~~Bottom--up approach}
\end{flushleft}
The procedures outlined above are used at two different stages
of our approach. First, we come back to the problem of fixing
all the parameters in the Higgs potential (2.10) at the scale
$M_Z$ in terms of the values taken for the set (2.1) and (2.2)
and use this procedure to obtain the couplings $\lambda_i(M_Z)$
from their boundary conditions (2.11) at $M_{SUSY}$.
In general $\lambda_i(M_Z)=\lambda_i(M_t,~M_{\tau},~M_{SUSY},
{}~\tan\beta,~g_k,~(M_X))$ as their RG evolution depends on the
Yukawa couplings. Given the values of $\lambda_i(M_Z)$ we obtain
the $\hat{m}^2_i(M_Z),~i=1,2,3$~, in terms of $M_Z,~\tan\beta$
and $M_A(M_Z)$ by minimizing the potential (2.10)\footnote{We
neglect the running of the pseudoscalar mass $M_A$, i.e. we work
in the approximation $M_A(M_Z)=M_A(M_A)$ which is correct to at
least few \mbox{GeV} accuracy.}. The relevant formulae can be found eg.
in ref.[13] and, for easy reference, are collected in the Appendix A.

The second stage of our approach consists in finding the soft
supersymmetry breaking
parameters $m_o,~M_o,~A_o,~B_o$  and the parameter $\mu_o$ at the $M_X$
which give radiatively induced spontaneous gauge symmetry breaking with
the chosen set of values for the low energy observables (2.1) and
(2.2). To this end, we notice that in {\it both regions}
$(M_Z,~M_{SUSY})^{[20]}$ and $(M_{SUSY},~M_X)^{[6]}$ the RG equations  for the
soft Higgs and gaugino mass parameters as well as for the
trilinear couplings $A_i$'s  can be schematically
written as follows $(t = \frac{1}{(4\pi)^2}\,\ln~\frac{M_X^2}{Q^2})$:
\begin{eqnarray}
&&\frac{d}{dt}~A_k = C^{ik}_{AA}~A_i + C^{ik}_{AM}~M_i\nonumber\\
&&\frac{d}{dt}~M_k = C^{ik}_{MM}~M_i\nonumber\\
&&\frac{d}{dt}~m^2_k = C^{ik}_{mm}~m^2_i+C^{ik}_{mM}~M^2_i+C^{ik}_{mA}~A^2_i+
C^k_{m\mu}~\mu^2\hspace*{5mm}k=1,2,Q,U\ldots\nonumber\\
&&\frac{d}{dt}~\mu = C_{\mu\mu}~\mu
\end{eqnarray}
(The equation for $\hat{m}^2_3$ will be discussed later. Note that the
parameter $B$
appears only in the equation for $\hat{m}^2_3$.). The coefficients
$C^{ik}$ in eqs.(2.12) depend on the gauge and Yukawa couplings and
change at $M_{SUSY}$ (below $M_{SUSY}$ they also depend on the
$\lambda_i$ couplings). Defining the multidimensional vector
\[
\bar{X}=(M_iM_j,~M_iA_j,~A_iA_j,~m^2_i,~\mu^2)
\]
(where eg. $A_iA_j$ denote all possible products of $A_i$'s
present in the model and similarly for the other quantities) and
using eqs. (2.12) one has
\begin{equation}
\frac{d}{dt}~\bar{X}(t) = \hat{C}(t)~\bar{X}(t)
\end{equation}
(where $\hat{C}(t)$ is a linear operator) i.e. we have  a set of
linear, homogeneous equations. Its solution reads:
\begin{eqnarray}
&&\bar{X}(t) = \hat{R}~(t,t_o)~\bar{X_o}~(t_o)\nonumber\\
\mbox{with} &&\hat{R}~(t,t_o) = T~exp\left\{\int^t_{t_o}\!\!\hat{C}
{}~(s)~ds\right\}
\end{eqnarray}
where $T$ means chronological ordering in case of non--commuting
operators $\hat{C}(t)$.
In view of the boundary conditions at $M_X$:
\[
A^o_i = A_o,~~~M^o_i = M_o,~~~m^o_i = m_o
\]
the solution (2.14) can be written more explicitly
as\footnote{Analogous equations have been discussed earlier in
ref.[14] with the coefficients derived analytically (in the case
of no sparticle decoupling and negligible $Y_b$).}(we recall that
$\hat{m}^2_i=m^2_i+\mu^2,~~i=1,2$).
\begin{equation}
\hat{m}^2_i(M_Z)=c_{ij}(M_t,~M_{\tau},~g^2_k,~\tan\beta,~M_{SUSY})~P^o_j,
{}~~~i = 1,2
\end{equation}
where
\[
P^o=(M^2_o,~m^2_o,~\mu^2_o,~A^2_o,~M_oA_o).
\]
Quite similarly one gets
\begin{equation}
m^2_{Q,U}(M_{SUSY})=b^j_{Q,U}(M_t,~M_{\tau},~g_k^2,~\tan\beta,~M_{SUSY})~P^o_j
\end{equation}

The coefficients $c_{ij}$ and $b^j_{Q,U}$ in the algebraic equations
(2.15) and (2.16) are found by integrating numerically
the RG equations with the boundary
conditions like $P^o_1=1,~~P^o_i=0$ for $i\neq 1$ etc. This is a
two-step procedure: first, the supersymmetric equations are integrated from
$M_X$ to $M_{SUSY}$ with those boundary conditions at $M_X$. Then the RG
equations with decoupled squarks and gluinos are integrated from
$M_{SUSY}$ to $M_Z$, with the boundary values at $M_{SUSY}$
obtained from the first integration. $M_{SUSY}$ is fixed in advance by the
choice of the squark masses, eq. (2.2).

Thus, we have a set of 4 algebraic equations for 4 independent
parameters $M_o,~m_o,~\mu_o,~A_o$. Finally, $B_o$ is fixed from
\[
\hat{m}^2_3(M_{SUSY}) = \mu(M_{SUSY})~B(M_{SUSY})
\]
by the equation
\begin{equation}
\hat{m}^2_3(M_{SUSY}) = d_1\mu_oM_o + d_2\mu_oA_o + d_3\mu_oB_o
\end{equation}
(as before, the coefficients $d_i$ are calculated from the RG equations).

At this point we can complete our discussion of implementing the
matching conditions for $\Delta m^2_i$ at $M_{SUSY}$. All the
terms (A.7b), (A.8b) and (A.9) are proportional to the
parameters $\mu$ and $A_i$ taken at $M_{SUSY}$. It follows from the
eqs. (2.12) that, analogously to (2.15),
\begin{eqnarray}
&&\mu = c\mu_o\nonumber\\
&&A_i = c^i_A~A_o + c^i_M~M_o\nonumber
\end{eqnarray}
and, therefore, the corrections $\Delta m^2_i$ have identical
structure in terms of the unification scale parameters as the
$m^2_i$'s themselves. So, their inclusion amounts to some
modification of the coefficients $c_{ij}$ and $b^j_{Q,U}$ in the eq. (2.15)
and (2.16).

Our procedure is now completed and can be summarized as follows:
for any chosen set of values for the physical parameters (2.2)
we find the corresponding low energy parameters of the effective
lagrangian and then solve the algebraic equations (2.15), (2.16)
and (2.17) to find the lagrangian parameters at $M_X$ which give those
chosen values of $\tan\beta,~M_A,~m^2_Q$ and $m^2_U$. We
note that for each set (2.2) there are up to 8 solutions for
the parameters at $M_X$: the sign of $m_o$ is taken to be
positive by convention (the lagrangian is symmetric under an
overall change of sign of all soft breaking terms) but the
solutions appear in pairs of $\pm M_o$ and $\pm\mu_o$ (the
solutions e.g. for $M_o<0$ are the same as the solutions for
$M_o>0$ and the reversed signs of all parameters, with exception
of $m_o$). The remaining two-fold ambiguity follows from the fact that the set
(2.15--2.16) is equivalent to a quadratic equation for the ratio
$A_o/M_o$. This overall eight-fold ambiguity is simply a
reflection of the fact that the set (2.2) is not a complete set of
observables in the model: different solutions give different
prediction for other observables.
Finally, one should stress that for a given set of values for
the observables (2.2) our equations may have no solution at all. This is, in
fact, the most often encountered case for a random choice of the
values (2.2) and it simply means that the given set cannot be
obtained in the model. Our procedure in this paper is to fix
$M_t$ and $\tan\beta$ and then to scan systematically the values
of $M_A,~ m^2_Q$ and $m^2_U$ in the range, say, from 0 to 1 TeV,
by varying them in steps of a few GeV in search for solutions to our
equations. Thus, for a given
choice of $M_t$ and $\tan\beta$ we get a range of values for
$M_A,~m^2_Q$ and $m^2_U$ which are obtainable in the model (i.e.
for which the solutions to eqs. (2.15) and (2.16) exist). For each solution we
can then calculate the values of the other physical observables.
Finally, we impose on the obtained solutions the "no fine
tuning" constraint and all the presently available experimental constraints.\\
\vspace*{5pt}
\begin{flushleft}
{\sc {\bf 3.~~Qualitative Features of the Solutions}}
\end{flushleft}
\renewcommand{\theequation}{3.\arabic{equation}}
\setcounter{equation}{0}

\indent In this section we discuss those properties of our
solutions which follow from the requirement of radiatively
induced gauge symmetry breaking alone. We also include the
constraint of all masses squared to be positive and the usual
one of no deeper lying minima of the scalar potential that break
charge and/or colour${[15]}$: $A^2_t<3(m^2_Q+m^2_U+\hat{m}^2_2)$ and similarly
for the remaining $A_i$'s at low energy, as well as the analogous condition
at the unification scale.  Finally, we demand that the
scalar potential should be bounded from below at  the scale $M_X$.

The "naturalness" constraint and the constraints from the present experimental
limits will be discussed in the next sections.
As an introduction to our results let us recall certain
properties of the low energy potential (2.10). One can verify that
radiative corrections  stabilize the potential$^{[6,14]}$.
So, the low energy mass
parameters $\hat{m}_i,~~i$=1,2,3,  which correspond to the chosen
values of $\tan\beta$ and $M_A$, do not need, in general, to
satisfy at $M_Z$ the well known stability condition for the
supersymmetric tree level potential: $\hat{m}^4_3~<~(\hat{m}^2_1
+ \hat{m}^2_2)^2/4$. Secondly, as already mentioned earlier, up
to the standard model 1--loop corrections, the
effective potential is now renormalization scale independent in
the range $(M_Z,~~M_{SUSY})$. This means that the gauge symmetry
is broken at the tree level already at $M_{SUSY}$,
i.e. at the scale where the quartic couplings in the potential
(2.10) recover their supersymmetric values.\footnote{Another way
of seeing this is to remember that the driving force for the
difference in the $m^2_1$ and $m^2_2$ RG evolution from the
scale $M_X$ (which gives the breaking at low energy) are squarks
and they are decoupled at $M_{SUSY}$.} Note, however, that the
individual parameters in the potential and in the lagrangian, in
general, run sizably in the range $M_Z,~M_{SUSY}$ (particularly
the quartic scalar couplings !). Physical quantities at the
scale $M_Z$ receive in addition the SM 1--loop corrections
(particularly, from the top quark and vector boson exchange)
which are included into the analysis by our choice of the
renormalization scale $Q=M_Z$.
\indent Next, it is instructive to analyze the coefficients in eqs.(2.15)
and (2.16) and to understand them qualitatively with help of the RG
equations. A sample of coefficients for $\hat{m}^2_i,~~m^2_Q$ and $m^2_U$
is given in Table 1. Of special interest is their dependence on
the top quark mass, $\tan\beta$ and their modification by
radiative corrections. A remark is in order here: it is the
usual procedure followed by many authors to solve the
supersymmetric RG analytically in some approximation. We could,
of course, follow the same approach here to calculate the
coefficients in eq. (2.15) and (2.16). However, the approximations
involved in this procedure make it very difficult, if not
impossible, to discuss the most interesting, just mentioned,
dependences. Therefore, we prefer to base our discussion on
numerical results, guided by general qualitative features which
can be inferred directly from the RG equations.
Let us first look at $\hat{m}^2_1$ and $\hat{m}^2_2$
for small $\tan\beta$. Taking into account the definitions
$\hat{m}^2_i=m^2_i+\mu^2$ and  the initial conditions $m^2_i
=m^2_o$, we expect from the RG equations (dominated by the terms
proportional to $Y_t$ -- see the Appendix B) that for $\hat{m}^2_1$
the coefficients are
close to 1 for $m^2_o$ and $\mu^2_o$, positive and smaller for
$M_o^2$ and negligible for $A^2_o$ and $M_oA_o$. For $\hat{m}^2_2$ the
coefficient of $\mu^2_o$ is expected again to be 0(1) but the
coefficient of $m^2_o$ gets additional negative contribution
from the terms proportional to $Y_t$ which largely cancels the
+1 contribution from the initial condition. The most interesting
is the coefficient of $M^2_o$ which is 0(1) and negative !
This looks contradictory to the positive gauge terms in the RG
equation for $m^2_2$ (the same as for $m^2_1$) and
is the reflection of the running of the scalar masses
in the equation for $m^2_2$. We have
\begin{eqnarray}
\frac{d}{dt}~\left( m^2_Q,~m^2_U\right)&&=~~ \frac{16}{3}~g^2_3M^2_3
+\ldots\nonumber\\
\frac{d}{dt}(M_3/g^2_3)&&=~~ 0,~~~~~~~~~~~~~~M_3(M_X) = M_o
\end{eqnarray}
and this strong dependence on $M^2_o$ is transmitted to $m^2_2$
through the negative term $-3Y_t(m^2_Q + m^2_U + \ldots)$
in the equation for $m^2_2$. Naively a second order effect, it is in
fact the dominant one due to the large logarithmic enhancements (and
the running of $M_3$) and it
plays an important role in understanding our results. In
particular it explains strong dependence of the $M^2_o$
coefficient in $m^2_2$ on $M_{SUSY}$: the larger is this scale the smaller is
the absolute value of the negative
$M^2_o$ coefficient, due to the earlier decoupling of the gluino term in
eq.(3.1). This behaviour has also important implications for the fine
tuning problem (to be discussed in the next section) before and
after radiative corrections. Indeed, in our formalism, neglecting
radiative corrections to the scalar potential means setting $M_{SUSY}=M_Z$
for any values of the
stop mass parameters and we see (the numbers in brackets in Table 1)
that the $M_o^2$ coefficients in
$m^2_2$ are then, for the same set of values for the variables (2.2), much
larger in the absolute values than after radiative corrections.

\indent It is also very important to recall that the low energy values
of $m^2_Q$ and $m^2_U$ are almost directly given by the values
of $M^2_o$. The next--to--leading dependence is on $m^2_o$ but
with the coefficients by factor O(10) smaller then for $M^2_o$.
Again, the $M_o^2$ coefficients decrease substantially after
radiative corrections.

\indent The $\tan\beta$ and $M_t$ dependence of the coefficients
is mainly determined by the behaviour of the Yukawa couplings
and their contribution to the RG equations. For fixed $M_t$ and
increasing $\tan\beta$ the $Y_b$ becomes larger and the
$Y_t$--smaller (eventually, $Y_t/Y_b\cong 1$ for $\tan\beta\cong
m_t(M_t)/m_b(M_t)$). Accordingly, the coefficients in the equation for
$\hat{m}^2_1$ behave as follows: the $M^2_o$ coefficient is
driven from positive to negative values, the one for the $m^2_o$ is also
decreasing but remains positive, due to the $m^2_o$ contribution
to $\hat{m}^2_1$, and the $M_oA_o$ coefficient is slowly rising
from zero to $\sim 0.4.$ This behaviour follows from the
increase of $Y_b$ which starts to play the same role for
$\hat{m}^2_1$ as $Y_t$ for $\hat{m}^2_2$. At the same time the
$Y_t$ is decreasing (but it changes much less than $Y_b$) and,
in consequence, the coefficients in the equation for
$\hat{m}^2_2$ change just in the opposite direction. Still, even
for $\tan\beta\approx m_t/m_b$ the coefficients
remain different enough from those in the expression for $\hat{m}^2_1$ to
assure the existence of solutions$^{[16]}$. The dependence on $M_t$ (for
fixed $\tan\beta$) can be easily understood in a similar way, and follows
from changing $Y_t$. (We note also the decrease of the
$\mu^2_o$ coefficients with increasing $Y_t$).

\indent Generically, the mass parameters of the low energy potential
satisfy the relation $\hat{m}^2_1\gg \hat{m}^2_2$ and $\hat{m}^2_2\leq 0$.
Thus,
to understand qualitatively the pattern of our solutions, in the
first approximation we can take $m^2_2=0$. In a large region of the
parameter space we then have:
\begin{eqnarray}
&&\hat{m}^2_1\approx a~M^2_o+b~m^2_o+c~\mu^2_o\nonumber\\
&&\hat{m}^2_2\approx -\alpha M^2_o+\beta m^2_o+\gamma\mu^2_o\approx 0
\end{eqnarray}
with $b,~c,~\alpha,~\gamma$ positive and $a$ and $\beta$ generally
smaller and of both signs. This structure, together with the
relation $M_A\approx m^2_1$ and
\begin{equation}
m^2_{\tilde{q}}\approx C_{\tilde{q}}\left( M^2_o+O\left(\frac{1}{10}
\right) m^2_o\right)
\end{equation}
explains the correlations seen in our solutions for the absolute
values of the pairs $(M_o,~\mu_o$), $(M_o,~m_o)$ and
$(\mu_o,~m_o)$. It is most visible in the $(M_o,~\mu_o)$
correlation (or, equivalently, $(M_2,~\mu$)) which is
interesting in the context of the
neutralino properties predicted by the model. It is clear from
the coefficients in Table 1 that, as long as we can neglect the
term $M_oA_o$ in the equation for $\hat{m}^2_2$ (and neglecting
also the other small coefficients), crudely speaking, the solutions exhibit
approximately linear correlation $M^2_o\approx (\gamma/\alpha)\mu^2_o$. The
actual allowed regions in the space $(M_2, \mu)$ are determined
by the integration over the soft scalar masses $m^2_Q$ and
$m^2_U$, and over the pseudoscalar mass $M_A$.
They are shown in Fig. 5 for $\tan\beta = 2$ and
$10$ for $M_t = 130~~\mbox{GeV}$ and for $\tan\beta=2$ and 30 for $M_t=180~~
\mbox{GeV}$.
A remark has to be made here.
In this section we refer to all the solutions shown
in the Figures. The distinction between experimentally excluded
or allowed and natural or unnatural solutions (which is marked
in the Figures) will be discussed in the subsequent sections. The dependence of
the $(M_2, \mu)$ correlation on $\tan\beta$ and $M_t$ follows
directly from the discussed above properties of the coefficients
and in the first, crude, approximation is just the reflection of
the changing slope $a$ in $M_o^2\approx a\mu^2_o$. The
correlation $(M_2, \mu)$ has direct consequence for the
predicted neutralino properties. It is well known$^{[17]}$, and follows
directly from the structure of the gaugino--higgsino mass matrix, that in
the limit of large $M_2\gg\mu$ the lightest neutralino becomes
almost pure higgsino with $M_n=|\mu|$ and in the limit of large
$\mu\gg M_2$ -- pure  gaugino with $M_n\approx
M_2/2$. The pattern observed in the $(M_2, \mu)$ plane explains
qualitatively the results shown in Fig. 7. We plot there the
gaugino content $Z_n=Z^2(1,1)+Z^2(1,2)$ of the lightest
neutralino as a function of the neutralino mass, as predicted by
our solutions. (As usual we define
$\chi_1=Z(1,1)\hat{B}+Z(1,2)\hat{W}^3+Z(1,3)
\hat{H}_1+Z(1,4)\hat{H}_2$ where $Z(i,j)$ is the real orthogonal
matrix that diagonalizes the neutralino mass matrix.)  In
particular, the $M_t$ and $\tan\beta$ dependence of those
results is easily interpretable in terms of the analogous
dependence of the $(M_2, \mu)$ correlation.

\indent The above simple discussion breaks down and a new
feature of our solutions becomes visible for the values of $M_t$
and $\tan\beta$ such that the coefficient of the $M^2_o$ term in
the equation for $\hat{m}^2_1$ is much smaller in its absolute value than
that of the $M_oA_o$ term. For instance, this is the case for
$M_t=160~~\mbox{GeV}$
and $\tan\beta=20$. Then the dependence on $M_o$ of the
$\hat{m}^2_1$ comes from the $M_oA_o$ term and is correlated
with the values of $A_o$. Positive $A_o$ allows for larger positive $M_o$,
since then the $M_oA_o$ term in $\hat{m}^2_2$ (no longer negligible) is
positive and helps to cancel out the negative $M_o^2$ term. Neutralinos with
large gaugino content (large $\mu$, small $M_2$) appear mainly
for light $m^2_Q$ (hence small $M_o$ and ($-|A|_o$) branch) and large
$M_A\approx\hat{m}^2_1$ (hence large $\mu$). For even larger
values of $M_t$ and $\tan\beta$ (e.g.
$M_t=180~\mbox{GeV},~\tan\beta=30$) the coefficient of the $M_o^2$ term
in the equation for $\hat{m}_1^2$ becomes again large in the
absolute value (but negative) and the effect disappears.

Another interesting correlation present in the model is seen in
the ($m_{\tilde{t}},~M_o$) plane ($m_{\tilde{t}}$ is the mass of the
lightest squark), as illustrated in Fig. 10. It follows from
the structure of eq. (3.3) and from the qualitative features of
the coefficients shown in Table 1: the $m_{\tilde{t}}$ is almost
directly related to $M_o$\footnote{Strinctly speaking, this is
true for the soft squark masses. However, the results obtained
after diagonalization of the stop mass matrix (presented in
Fig. 10), with the mixing terms predicted by our solutions, show
very similar pattern.}. One can also expect some correlation in
the ($m_{\tilde{t}},~M_A$) plane. As explained above, $M_o$ is
correlated with $\mu_o$ (up to the contribution of the $m_o$ and
$A_o$ terms) and therefore $M_A^2\cong \hat{m}_1^2$ should be
correlated with $m_{\tilde{t}}$. However, as seen from Fig. 12,
the dependence on the additional parameters makes the final
correlation rather weak. An obvious correlation exists between
$m_{\tau}$ and the $m_o$, Fig. 11. The qualitative difference between
Fig. 11A, C and Fig. 11 B, D is due to the negative contribution
of the $Y_b$ term to the RG equation for $m_{\tilde{\tau}}$,
which is increasing for large $\tan\beta$.

\indent In Fig.1--4 we also show our solutions in the planes $(m_o,
M_o)$, $(m_o,~~A_o)$, $(A_o,~~M_o)$ and ($A_o/m_o,~~B_o/m_o$).
Clear positive correlation is observed in the $(m_o,A_o)$
projection. Interesting structure is seen in the ($A_o/m_o,~~B_o/m_o$) plane:
there is an almost linear correlation (see eq. (2.17) and Table 1) and two
branches of solutions exist, corresponding to the two solution
with $\mu=\pm|\mu|$, and with the gap depending on the value of
$\hat{m}^2_3$. Only weak correlation is observed in the $(m_o, M_o)$ plane.
Obviously, it
will be very interesting to study correlations in 3 and more variables.

\indent So far we have been discussing the qualitative features of the
predictions based on the solutions to eqs.(2.15), (2.16) and (2.17), i.e.
following from the requirement of radiatively induced gauge symmetry
breaking. In the next two sections we impose two additional
constraints: no fine tuning and the present experimental limits.\\
\vspace*{5pt}
\begin{flushleft}
{\sc {\bf 4.~~Fine Tuning}}
\end{flushleft}
\renewcommand{\theequation}{4.\arabic{equation}}
\setcounter{equation}{0}

\indent The basic feature of the model is to predict the low
energy parameters in terms of the unification scale parameters
which, hopefully, can be obtained in future from some underlying
theory. This procedure involves extrapolations of many orders of
magnitude in the energy scale. In this context it is important to
address the question of the degree of fine tuning of the
parameters at $M_X$ which is necessary to get the correct low
energy results. Precise values of the input parameters do not,
by themselves, mean fine tuning yet. Only when the precision we need
for the input parameters is much higher than the precision we
require for the low energy results, it means fine tuning. In other words
it is the question of response in the low energy results $Q_i$ to small
changes in the high energy input $P_j$, which is measured by$^{[2]}$
\begin{equation}
\Delta_{ij} = \frac{\delta Q_i}{Q_i}~/~\frac{\delta P_j}{P_i}
\end{equation}
Equivalently this is also a measure of cancellations in
equations like (2.15): for instance, calculating the mass
parameters $\hat{m}^2_i$ in the low energy lagrangian in terms
of the high energy parameters $P_i$ (using the coefficients like
those in Table 1) the single terms on the rhs should not be much
larger than the calculated masses. To some extent it is a matter
of taste how big cancellations (i.e. how big $\Delta_{ij}$'s) are
still tolerable. The coefficients $c_{ij}$ in eqs. (2.15)--(2.17) depend
on large logaritms $\ln M_X/M_Z\cong 0(30)$, so we take here the
attitude that cancellations of this order of magnitude are still
"natural".

\indent Our actual interest is to avoid fine tuning when calculating the
physical quantities like $M_Z$ (or $v$), $\tan\beta, M_t$, etc.
So we look for solutions with acceptable all the following derivatives
\begin{eqnarray}
\Delta_{ij}=\frac{P_i}{Q_j}~~\frac{\partial Q_j}{\partial
P_i}~~,&&Q_i\equiv\left\{
Y_t,Y_{\tau},v,\tan\beta,M_A,m_{Q},m_U\right\}\nonumber\\
&&P_i\equiv\left\{ Y^o_t,Y^o_{\tau},M_o,m_o,\mu_o,A_o,B_o\right\}
\end{eqnarray}

We shall consider a solution to be "natural" if
\begin{equation}
\mid\Delta_{ij}\mid~~<~A~,~~~ \mbox{for}~~i,j=1,\ldots 7
\end{equation}
where, following the arguments given above, we sort the
solutions according to $A=10,~30$ and no constraint on
$|\Delta_{ij}|$. It turns out, however, that in the range of the
low energy parameters studied in this paper,
$M_A,~m_Q,~m_U~<~1~TeV$ almost all  solutions satisfy $|\Delta_{ij}|<100$
(for moderate values of $\tan\beta$ discussed here).

For given values of $v$ and $\tan\beta$ the constraints (4.3)
mean upper limits on the unification scale parameters,
or equivalently, upper limits on $M_A$ and the stop
masses for which "natural" solutions exists. In general, it
is difficult to avoid large cancellations simultaneously in all
the low energy mass parameters.

There are two main effects which ease the problem of fine tuning
after radiative corrections. These are: increase of the absolute
value of $\hat{m}^2_2$ for the same set of low energy parameters (see eq. (A.4)
and remember that, generically $\hat{m}^2_2\leq 0$) and (the discussed
earlier) decrease of some coefficients in
eqs.(2.15)--(2.17), particularly the ones of the $M^2_o$ terms. The overall
effect is that the maximal acceptable $M_2$
considered as a function of the max $|\Delta_{ij}|$ is,
with radiative corrections included, almost twice as large
as without. Similar effect exists for
$m_{\tilde{t}}$ and $\mu$. This results in a sizable enlarging of the
"natural" regions (for any chosen value of $A$) on all the
Figures. For the illustration of this point we mark in Figs, 1, 5,
7, 12 (by the solid contours) the "natural" regions, for $A=30$,
obtained without radiative corrections. One should also remark that for
$A=10$ the allowed parameter space of the model is small even with
radiative corrections included. The fine tuning
problem becomes more severe for $\tan\beta$ close to 1 and to
$m_t/m_b$, i.e. close to the limits of the $\tan\beta$ range
obtainable in this model. Solutions for the large $\tan\beta$ case
will be discussed in more detail in Section 6.

A comment is in order here on our method of calculating derivatives
$\partial Q_i/\partial P_j$ with $P_k$ fixed, for $k\neq j$.
Note that the coefficients $c_{ij}$ in eqs.(2.15)--(2.16) are fixed for
given values of the low energy parameters (in particular they
depend on the decoupling point which is fixed in terms of the low
energy squark mass parameters). Thus, it is straightforward for
us to calculate the inverse matrix  $\partial
P_j/\partial Q_i, Q_k$  fixed for, $k\neq i$,
and to find by inversion the matrix of interest. The direct
calculation of the latter is in our approach cumbersome because fixing
the unification scale parameters rather then the low energy parameters
makes the coefficient $c_{ij}$ dependent on them (the decoupling
point depends then on $P_i$'s). Thus, we calculate:
\[
\left.\frac{\partial Q_i}{\partial P_j}\right |_{P_{i\neq
j}} =~~\left[\left.\frac{\partial P_k}{\partial Q_l}\right
|_{Q_{m\neq l}}~\right]^{-1}
\]
Note, that to get the correct answer for the lhs derivatives one has
to calculate the complete matrix on the rhs. Most often, the largest
derivatives are those of the $v$ with respect to $Y_t,~\mu_o$ and $A_o$
and of $\tan\beta$ with respect to $B_o,~m_o,~A_o$ and $Y_t$.
Finally, one should remark that in eq. (4.2) we consider only
the parameters which are used in the calculation as independent
input parameters. It also happens that the derivatives of some
other physical quantities (e.g. sparticle masses), taken with
respect to the high energy set $P_i$, become large, even when
$\Delta_{ij}$'s defined in eq. (4.2) are small.
\newpage
\noindent {\sc {\bf 5.~~Experimental Constraints and Final Results}}
\renewcommand{\theequation}{5.\arabic{equation}}
\setcounter{equation}{0}
\vspace*{5pt}\\
Let us recapitulate our procedure: For some chosen values of the
low energy physical parameters $M_t,~\tan\beta$, $M_A$,
$m_Q$ and $m_U$ (see eq. (2.2)) we look for solutions to eqs. (2.15)--
(2.17). These are the equations for the unification scale parameters,
derived from the requirement of  radiatively induced gauge
symmetry breaking in the minimal supergravity model. The
existence (non--existence) of solutions for a given set of input
values means that this particular set can (cannot) be realized
in the model. In this paper we present results for a few generic
top mass and $\tan\beta$ values: $M_t = 130~~\mbox{GeV}$ and $\tan\beta
= 2$ and $10$; $M_t = 180~~\mbox{GeV}$ and $\tan\beta = 2$ and $30$. In
the next Section we discuss very large $\tan\beta$ cases:
$\tan\beta\approx m_t/m_b$. In each case the $M_A,~m_Q$ and $m_U$ have been
scanned in the region from 0 to $1~\mbox{TeV}$ in steps of a few $\mbox{GeV}$.
Once some solutions for the unification scale parameters are found we can
predict the values of any other low energy observables,
corresponding in the model to our set of input data. We present
our results in  form of the scatter plots obtained by our
scanning procedure. In this paper we limit ourselves to
two--dimensional scatter plots, but
multi--dimensional correlations are also of interest. In
Fig. 1--4 we present  our solutions in the unification scale variables
$(m_o,~M_o),~~(m_o,~A_o),~~(A_o,~M_o)$ and $(B_o/m_o,~~A_o/m_o)$. Projections
onto the variables $(M_2,~\mu),~~(M_n,~M_A),~~(Z_n,~M_n),~~(M_c,~M_A),
{}~~(M_c,~M_n),~~(m_{\tilde{t}},~M_o),~~(m_{\tilde{\tau}},~m_o)$,
$(m_{\tilde{t}},~M_A)$ and $(M_c,~m_{\tilde{t}})$ are shown in Fig. 5--13.
(We recall the notation: $M_o,~m_o,~\mu_o,\\A_o,~B_o$ are the high scale
parameters; $M_2$ is the low energy $SU(2)$
gaugino mass, $\mu$ is the low energy supersymmetric Higgs mixing parameter;
$M_A,~M_n,~M_c,~m_{\tilde{t}},~m_{\tilde{\tau}}$~ are the pseudoscalar, and
the lightest neutralino, chargino, stop and stau physical masses,
respectively, and $Z_n$ is the gaugino content of the lightest neutralino.
Three different classes of solutions are introduced: a) solutions already
excluded by the present experimental constraints, b) allowed by
the present experiments and satisfying the criterion $\max
|\Delta_{ij}|<~30$, c) allowed by experiment and
with $30~<$ $\max |\Delta_{ij}|<~100$. (In the considered region of
$M_A,~m_Q,~m_U<1~TeV$ and for those moderate values of $\tan\beta$, solutions
with $\max |\Delta_{ij}|>~100$ occur quite rarely.) It is worth
pointing out that, in our 2--dimensional projections of the
5--dimensional parameter space, those three different classes of
solutions are often placed in overlapping regions. For clarity of the
Figures, only part of the obtained solutions is plotted: only as
many of them as
is needed to properly mark various domains in the parameter space.
Additional important  information given in some of the
Figures is the sensitivity of our results to  radiative
corrections. The solid contours in  Fig. 1,5,7,12  encircle the
regions where the class (b) solutions are present,  with no
radiative corrections included. In the same Figures, the
dashed contours show the regions where $\max |\Delta_{ij}|<10$.

For the purpose of this paper we have taken the following
"experimental" limits as the constraints for our solutions:
\begin{eqnarray}
&&m_{\tilde{g}}>120~\mbox{GeV},~m_{\tilde{l}}>45~\mbox{GeV},~M_c>45~\mbox{GeV}\nonumber\\
&&|Z^2(1,3)-Z^2(1,4)|<0.03,~|Z(1,3)~Z(2,3)-Z(1,4)~Z(2,4)|<0.04\nonumber\\
&&(\mbox{for}~ M_n<45~\mbox{GeV})~~~~~~~~~~~~(\mbox{for}~ M_{n_1}+M_{n_2}<45~
\mbox{GeV})\nonumber\\
&&m_{\tilde{q}}>100~\mbox{GeV}~~(\mbox{for the first two generations of
squarks),}
\nonumber\\
&&m_{\tilde{t}}>45~\mbox{GeV},~~M_A>40~\mbox{GeV}.
\end{eqnarray}
Some of the above limits are somewhat rough representation of the
actual experimental data. In particular, strictly speaking, at present there
is no absolute lower limit on the stop mass$^{[18]}$. The bounds
from $p\bar{p}$ colliders do not apply to it; those bounds
assume degenerate squarks and this assumption fails for strong
L--R mixing of the $\tilde{t}$--squarks. Also searches at the $e^+e^-$
colliders are affected since the L--R mixing can reduce the $Z\tilde{t}_1
\tilde{t}_1$ coupling\footnote{Similar remark applies also to stau
leptons.}. In the model considered, each solution gives definite
predictions for mass mixings and couplings and in principle a
more detailed phenomenological analysis with respect to the
constraints from the present $p\bar{p}$ and $e^+e^-$ data is
possible for each solution. We postpone it, however, for future
study and take (5.1) as sufficient for our discussion here.

As it has already been mentioned several times,  the model shows a
high degree of correlation among various physical quantities
which results from the small number of free parameters. We have
discussed in  Section 3 that a number of correlations follows
from the requirement of radiatively induced gauge symmetry
breaking. We can add now to this discussion some remarks
following from the fine tuning and the experimental constraints.
As expected, the larger the values of the high energy parameters
the more fine tuning is required (see Fig. 1--4). For instance, for
$\max |\Delta_{ij}|<30$ we get $M_o<(200-400)~\mbox{GeV}$ and $m_o<
(400-900)~\mbox{GeV}$, depending on the values of the top quark mass
and $\tan\beta$. Simultaneously, the ratio $m_o/M_o$ can vary,
approximately, in the range (0.1--10). As we know from Section
3, the gauge symmetry cannot be broken for $m_o\approx
M_o\approx 0$. At least one of them must be visibly different
from zero. The experimental constraints put somewhat stronger
lower bound on $M_o(>70~\mbox{GeV})$ than on $m_o(>(20-30)~\mbox{GeV})$
and acceptable solutions with $m_o\approx A_o\approx B_o\approx 0$ exist.

At the quantitative level, the predictions of the model show
strong dependence on the top quark mass, $\tan\beta$ and on the
inclusion of the radiative corrections. For a detailed survey of
the predictions for the physical (low energy) observables we
refer the reader to Fig. 5--12. Here we would like to stress
again the interesting correlations between $M_2$ and $\mu$, between $M_t$
and $Z_n$ (the gaugino content of the lightest neutralino), $\tan\beta$ and
$Z_n$, $Z_n$ and $M_n,~~m_{\tilde{t}}$ and $M_A$ etc.
Similarly to the requirement of radiative gauge symmetry breaking, the
experimental constraints (5.1) have "global"
effects, i.e. each individual constraint affects not only this
concrete quantity but also all the others through the rigid
structure of the model. An interesting example is the interplay
of the experimental lower bounds on the gluino, chargino and
neutralino masses. In the model the three masses are
interconnected by the parameters $M_2$ (or $M_o$), $\mu$ and
$\tan\beta$ and it varies from case to case which experimental
bound is most operative in fixing e.g. the effective lower
bound for the neutralino mass. For neutralinos with large
higgsino content ($\mu$ small, $M_2$ large), i.e. with large
coupling to the $Z$, the LEP data give directly $M_n>45~\mbox{GeV}$.
For gaugino--like neutralinos ($M_2$ small, $\mu$ large) the
bound on $M_n\approx M_2/2$ follows from the bound on the gluino
mass ($M_2\approx 0.3\,m_{\tilde{g}}>40~\mbox{GeV}$) or from the LEP
bound on the chargino mass (for $M_2\ll \mu$ and $\mu>M_W$ one
has $M_c\sim M_2$). More precise numbers follow from the well
known complete chargino and neutralino mass matrices and depend
on the actual values of the parameters in the given case. As seen
e.g. in Fig. 7, the lower bound for $M_n$ is indeed in the range
(20--50) GeV, depending on $M_t$ and $\tan\beta$.

Another interesting example is the effective lower limit for the
pseudoscalar mass $M_A$. We have taken $M_A>40~\mbox{GeV}$ as the
experimental lower limit for $M_A$ but, as seen e.g. in Fig. 6,8
and 12, the effective bound (which follows from the other
constraints (5.1)) is much higher,
$M_A>175,~125,~205,~150~\mbox{GeV}$ for $(M_t,~\tan\beta)=(130,~2),~(130,~10),
{}~(180,~2)~\mbox{and}~(180,~30)$, respectively.  One can
understand it as follows: $M^2_A\approx \hat{m}^2_1$ and, for the
discussed here moderate values of $\tan\beta$, all coefficients
in the equation (2.15) for $\hat{m}^2_1$ are positive and large (see
Table 1). Therefore, small $M^2_A$ implies small values of
$M^2_o,~~m^2_o$ and $\mu^2_o$. However, as discussed above,
there are lower bounds on those parameters from the experimental
lower bounds on the gluino, neutralino and chargino masses
which, as shown in Fig. 6 and 8, give the effective lower bound
for $M_A$ of the order O(200 GeV).

For larger values of $M_t$ and $\tan\beta$ (e.g. $M_t=180~\mbox{GeV}$ and
$\tan\beta=
30$) the coefficients in $\hat{m}_1^2$ of the $M_o^2$ and
$\mu^2_o$ terms become very close to each other and of opposite
signs. The above argument does not then work any more:
arbitrarily large values of $M^2_o\approx\mu_o^2$ can give small
$M_A$. However, we see that, in the case considered, the
coefficients of the $M_o^2$ and $\mu^2_o$ terms in the
$\hat{m}_2^2$ are very different and, since
$\hat{m}^2_2\approx0$, to obtain radiatively induced gauge
symmetry breaking we need $\mu^2_o\gg M^2_o$ (as seen in Fig. 5).
Hence, the terms $M^2_o$ and $\mu^2_o$ in the $\hat{m}^2_1$
cannot cancel each other and we again get the effective lower
limit on $M_A$ higher then the present experimental lower limit.
Also note that similar arguments explain the correlation between
light neutralinos, with large gaugino content, and heavy
pseudoscalars which is observed for $M_t=180~\mbox{GeV}$.

For even larger values of $\tan\beta$ the situation again looks
differently: as will be discussed in more detail in the next
section, for $\tan\beta\approx m_t/m_b$ {\it all} the coefficients in the
expression for $\hat{m}_1^2$ are very close to the coefficients
for $\hat{m}^2_2$ and a light pseudoscalar $A$ becomes
compatible both with radiative symmetry breaking and with the
experimental constraints.

In general, we recall that the relative magnitude of the
coefficients of the $M^2_o$ and $\mu^2_o$ terms in the
expressions for $\hat{m}^2_1$ and $\hat{m}^2_2$ is responsible
for most of the structure which follows from the mechanism of
the symmetry breaking alone. The lower limit on $M_A$ is an
example of its interplay with experimental constraints, for
changing values of $M_t$ and $\tan\beta$.

In Table 2 we present several examples of the generic sparticle
spectra obtained in this model. In Table 3 we also give results obtained
with and without inclusion of radiative corrections for the same sets of the
input
parameters. We see that the difference between the two sets can be very
important.
\vspace*{5pt}
\begin{flushleft}
{\sc {\bf 6. Solutions for Large}} $\mbox{\boldmath $\tan\beta$}$
\end{flushleft}

\indent It has been discussed for some time$^{[21]}$ that the
observed pattern of the quark masses, namely strong isotopic
spin symmetry breaking in the third generation electroweak
doublet, can be understood as the property of the vacuum. In
models with two (or more Higgs doublets) such that two different
$vevs$ $v_1$ and $v_2$ are driving the down -- and up -- quark
masses, it is conceivable that $\tan\beta\equiv v_2/v_1\cong
m_t/m_b$, with the top and bottom Yukawa couplings (almost)
equal. Such a scenario is particularly interesting in the
context of supersymmetric grand unified models (they require at
least two Higgs doublets in the low energy effective lagrangian)
which, if based on the groups $SO(10)$ or $E_6$ with simple
Higgs structure, predict $Y_t=Y_b$ at the unification scale.
This scenario has recently been strongly advocated in a detailed
study of the fermion mass and mixing textures$^{[22]}$.

An interesting question is whether the minimum of the potential
with large $\tan\beta$ can be obtained by radiative gauge
symmetry breaking.
The answer to this question is indeed positive, as shown for the
first time in ref. [16] and subsequently investigated in a
number of papers$^{[19]}$.

Our bottom to top approach is particularly suitable for a
systematic search for large $\tan\beta$ solutions and a study of
their low energy properties. A sample of such solutions is shown
in Fig. 14--17 for $M_t=130~\mbox{GeV},~~\tan\beta=30$ and 31, and
$M_t=180~\mbox{GeV},~~ \tan\beta=50$ and 51. The most important points
about the large $\tan\beta$ solutions are the following: The
degree of fine tuning is rapidly increasing when we approach the
value of $\tan\beta$ such that $Y^o_t=Y^o_b$. For instance, for
$M_t=130~\mbox{GeV}$ and $\tan\beta=31$ (the corresponding value of
the $Y_t^o/Y^o_b$ at $M_X$ is 1.04) we have
not found any solutions with $\max |\Delta_{ij}|<100$ whereas for
$\tan = 30$ ($Y^o_t/Y^o_b\approx 1.08)$ there exist
solutions with $\max |\Delta_{ij}|<30$. This is easily
understandable from inspection of the RG equations: the masses
$m^2_1$ and $m^2_2$ run now very similarly and in consequence
the coefficients for $\hat{m}^2_1$ and $\hat{m}^2_2$ in eqs. (2.15) are very
close to each other. As an example, for
$M_t=130~\mbox{GeV},~~\tan\beta=31$ and $M_{SUSY}\cong 460~\mbox{GeV}$ we have:
\begin{eqnarray*}
&&\hat{m}^2_1=-0.95\,M^2_o-0.27\,A_o^2+0.41\,m^2_o+1.20\,\mu^2+0.47\,M_oA_o\\
&&\hat{m}^2_2=-1.07\,M^2_o-0.10\,A_o^2+0.45\,m_o^2+1.20\,\mu^2+0.39\,M_oA_o\\
&&\hat{m}^2_3=0.18\,\mu M_o-0.41\,\mu_oA_o+1.08\,\mu_oB_o
\end{eqnarray*}
\noindent At the same time, the input values of $\hat{m}_i^2$'s at
$M_Z$ are: $\hat{m}^2_1\approx (6-9)\times 10^3~\mbox{GeV}^2,~~\hat{m}^2_2
\approx -5\times 10^3~\mbox{GeV}^2$ and $\hat{m}^2_3\approx 70-160~
\mbox{GeV}^2$.
Clearly, we need large values of the parameters at $M_X$ (in
particular of $M_o$)\footnote{The small differences in the
coefficients are driven by the residual difference in the Yukawa
coupling and also by the different $U(1)$ charge assignement for
the right--handed up- and down- squarks. The latter (and the
slepton contributions) are sufficient for the existence of solutions even for
$Y_t^o=Y^o_b$ but at the expense of very large values of $M_o$ and
$\max |\Delta_{ij}|\sim\mbox{O}(10^3)$.} and a sizable amount of fine tuning
to solve those equations.

The above example illustrates several other points typical for
the large $\tan\beta$ scenario. These are: $\hat{m}^2_3\approx 0$ (see
eq. (A.5) and small pseudoscalar mass $M^2_A\approx
\hat{m}^2_1+\hat{m}^2_2$. The acceptable solutions exist now for small
values of $M_A$, since for large $\tan\beta$ all the coefficients in the
equations for $\hat{m}^2_1$ and $\hat{m}^2_2$ are close to each other and
moreover, in the combination $M^2_A=\hat{m}^2_1+\hat{m}^2_2$ the terms
$M^2_o$ and $\mu^2_o$ tend to cancel each other. Therefore, the arguments of
the previous Section do not apply: large $\tan\beta$ solutions in general
give light pseudoscalar Higgs bosons and a much heavier spectrum for
squarks and gluinos.

Another point is that $\hat{m}^2_3\approx 0$ can be realized in two
different ways: small $\mu_o$ or cancellations between the three
terms. It is the latter possibility which is chosen by most of
the solutions.

Finally, we stress that large $\tan\beta$ solutions are
particularly sensitive to radiative corrections included in our calculation.

In conclusion, large $\tan\beta$ scenario is compatible with
radiative gauge symmetry breaking, although the exact equality
$Y^o_t=Y^o_b$ can be achieved only at the expense of very strong
fine tuning. On the other hand, allowing for, say, (5--10)\%
difference in the two Yukawa couplings at $M_X$ brings the
necessary degree of fine tuning into  acceptable range. It
is quite possible that such difference is generated by high
energy thresholds and contributions from higher dimension
operators, but their impact on the Yukawa coupling unification has not
been investigated so far.\\
\vspace*{5pt}
\begin{flushleft}
{\sc {\bf 7.~~Conclusions}}
\end{flushleft}

\indent In this paper we have proposed a bottom--up approach to unified
supergravity models which give the minimal supersymmetric
standard model as the effective low energy theory. In this
approach the supergravity models are parametrized in terms of a
number of low energy physical parameters. The minimal model,
with universal boundary conditions at the unification scale and
five free mass parameters, has been studied in detail. With more
than five low energy input parameters, our set of equations
can be used to investigate non--minimal models, which do not
assume the universal boundary conditions at $M_X$. This will be
particularly interesting once some sparticles are discovered and
their masses known.

Even in the minimal model, there are several virtues of the
bottom--up approach. One is that the number of free parameters
is reduced to four (plus the top quark mass)
since the mass $M_Z$ is known. Secondly, a systematic and
complete investigation of the low energy predictions of the
model can be easily performed, by scanning the space of four
unknown {\it low energy} input parameters. Finally, with a
suitable choice of the input parameters, our approach is most
convenient for inclusion of radiative corrections generated by
large Yukawa couplings and particle--sparticle mass splitting:
by choosing the soft stop masses as the input parameters the
dominant 1--loop radiative corrections are precisely known
directly in terms of the input parameters.

The complete inclusion of radiative corrections generated by
large Yukawa couplings and particle--sparticle mass splitting is
the other main point of this work. They are found to have strong
effects on the degree of fine tuning needed in the model and on
the final results.

In the above framework, low energy phenomenology of the minimal
model has been explored in some detail. In particular, we have
imposed all the presently available experimental constraints on
the predictions of the model and obtained complete information
on those regions of the low energy physical observables which
are interesting for future experimental exploration.

Due to the very rigid structure of the minimal model, both the
requirement of radiative gauge symmetry breaking and the
experimental constraints have strong "global" effects: they
introduce a lot of correlations between different quantities
predicted by the model, as discussed in the previous sections.
Particularly interesting and illuminating is the dependence of
the predictions on the values of the top quark mass and
$\tan\beta$. In this respect, our result based on numerical
solutions to the RG equations show a very rich structure, which
is difficult to see fully in appproximate analytic calculations.
On the list of interesting results are, among others, neutralino
as the lightest neutral sparticle, the gaugino content of the
lightest neutralino, the lower limit for the pseudoscalar mass
$M_A$ and, in general, correlations among masses (and couplings)
of sparticles. The model predicts a rich sparticle spectrum to
be accesible at the next generation of colliders.

In the subsequent paper$^{[23]}$ we calculate in the same approach
the relic abundance of the lightest neutralino and impose the
additional constraint $\Omega h^2<1$ on the predictions of the model.

As a further extension of the present work, it will also be very
interesting to explore systematically other constraints which
may follow from more specific assumptions about physics at the
grand unification scale, such as e.g. proton life time
constraints in the supersymmetric $SU(5)$ model$^{[24]}$.\\
\vspace*{5pt}
\begin{flushleft}
{\sc Acknowledgments}
\end{flushleft}
It is a pleasure to acknowledge useful discussions with I.
Antoniadis, C. Bachas, P. Binetruy, H.P. Nilles, C. Savoy and S.
Raby.
\newpage
\begin{center}
{\sc {\bf APPENDIX A}}
\end{center}

\indent The minimization of the potential (2.10) gives
$$\tan^2\beta(\hat{m}^2_2+\lambda_2v^2)=\hat{m}^2_1+\lambda_2v^2+
(\lambda_1-\lambda_2)v^2_1$$
$$2\hat{m}^2_3=\left[\hat{m}^2_1+\hat{m}^2_2+\lambda_1v^2_1+\lambda_2v^2_2
+\left(\lambda_3+\lambda_4\right)v^2\right]\sin 2\beta
\eqno \mbox{(A.1)} $$
where $\tan\beta=v_2/v_1$ and $v^2=v^2_1+v^2_2$.\\
The pseudoscalar mass $M_A$ reads:
$$M^2_A=\hat{m}^2_1+\hat{m}^2_2+\lambda_1v^2_1+\lambda_2v^2_2+
(\lambda_3+\lambda_4)v^2$$
$$\mbox{}\hspace*{-7em}\approx\hat{m}^2_1+\hat{m}^2_2+\Delta\lambda_2v^2_2
\eqno \mbox{(A.2)} $$
The last relation holds in the approximation in which we neglect
all but the dominant radiative correction, from the top--stop
mass splitting, to $\lambda_2$. For $\hat{m}^2_i$'s we get the
following relations in terms of the input parameters
$M_A,~~\tan\beta$ and $v^2$:
$$\mbox{}\hspace*{-5.5em}\hat{m}^2_1=M^2_A\sin^2\beta+\lambda_1v^2\frac{\tan^2
\beta-1}{\tan^2\beta+1}-(\lambda_1+\lambda_3+\lambda_4)v^2_2$$
$$\mbox{}\hspace*{-11.5em}\approx M^2_A\sin^2\beta+{1\over{2}}M^2_Z
\frac{\tan^2\beta-1}{\tan^2\beta+1}
\eqno \mbox{(A.3)} $$
{}\vspace*{10pt}\\
$$\mbox{}\hspace*{-5em}\hat{m}^2_2=M^2_A\cos^2\beta-\lambda_2v^2\frac{\tan^2
\beta}{1+\tan^2\beta}-\cos^2\beta(\lambda_3+\lambda_4)v^2$$
$$\mbox{}\hspace*{-4em}\approx M^2_A\cos^2\beta-{1\over{2}}M^2_Z\frac
{\tan^2\beta-1}{\tan^2\beta+1}-\Delta\lambda_2v^2\frac{\tan^2\beta}{1+\tan^2
\beta}\eqno \mbox{(A.4)} $$
{}\vspace*{10pt}\\
$$\mbox{}\hspace*{-21em}\hat{m}^2_3={1\over{2}}M^2_A\sin 2\beta\eqno
\mbox{(A.5)} $$

\indent 1--loop corrections from the squark loops to the
parameters of the potential (2.10) (calculated from Feynman
diagrams in the "symmetric" phase of the model$^{[6]}$) are as follows:
$$\hat{m}^2_i(eff)=\hat{m}^2_i+\Delta m^2_i,\hspace*{5mm}i=1,2,3\eqno
\mbox{(A.6)} $$
\noindent where
$$\hspace*{-3em}(4\pi)^2\Delta m^2_1=3\sum_A\left[
{1\over{6}}g^2_1d_A+Y^{A^2}_d\left(a_o
\left( m^2_{Q_A}\right) +a_o\left( m^2_{D_A}\right)\right)\right.\eqno
\mbox{(A.7a)} $$
$$\mbox{}\hspace*{6em}\left.+\,\mu^2Y^{A^2}_uf\left( m^2_{Q_A}, m^2_{U_A}
\right)+A^2_{D_A}Y^{A^2}_df\left(m^2_{Q_A}, m^2_{D_A}\right)\right]
\eqno \mbox{(A.7b)} $$
\vspace*{10pt}\\
$$\mbox{}\hspace*{-3em}(4\pi)^2\Delta m^2_2=3\sum_A\left[ -\,{1\over{6}}g^2_1
d_A+Y^{A^2}_u\left(a_o\left( m^2_{Q_A}\right) +a_o\left( m^2_{U_A}\right)
\right)\right.\eqno \mbox{(A.8a)} $$
$$\hspace*{5em}\left.+\,\mu^2Y^{A^2}_df\left( m^2_{Q_A}, m^2_{D_A}
\right)+A^2_{U_A}Y^{A^2}_uf\left( m^2_{Q_A}, m^2_{U_A}\right)\right]
\eqno \mbox{(A.8b)} $$
\vspace*{10pt}\\
$$\mbox{}\hspace*{-8.5em}(4\pi)^2\Delta m^2_3={3\over{2}}\,\mu\sum_A\left(
A_{U_A}Y^{A^2}_uf\left(
m^2_{Q_A},m^2_{U_A}\right)\right.$$
$$\mbox{}\hspace*{-7.5em}\left.+\,A_{D_A}Y^{A^2}_df\left(m^2_{Q_A},m^2_{D_A}
\right)\right)\eqno \mbox{(A.9)} $$
where
$$f\left( m^2_1, m^2_2\right) =\,-1+\,\frac{m^2_1}{m^2_1-m^2_2}\,\log\,
\frac{m^2_1}{\Lambda^2}\,+\,\frac{m^2_2}{m^2_2-m^2_1}\,\log\,\frac{m^2_2}
{\Lambda^2}\eqno \mbox{(A.10)} $$
\vspace*{10pt}\\
$$d_A\equiv 2\,a_o\left( m^2_{U_A}\right)-a_o\left( m^2_{D_A}\right)-a_o
\left( m^2_{Q_A}\right)\eqno \mbox{(A.11)} $$
and\\
\vspace*{10pt}
$$a_o\left( m^2\right) =(4\pi)^2\int\!\!\frac{d^d\,k}{(2\pi)^d}\,\frac{i}
{k^2-m^2}=m^2\left(\eta+\log\,\frac{m^2}{\Lambda^2}\,-1\right)$$
$$\eta = \frac{2}{d-4}\,+\gamma_E-\log 4\pi\eqno \mbox{(A.12)} $$
\vspace*{25pt}\\
We recall the notation: $\hat{m}_i=m^2_i+\mu^2,~~i=1,2$, where $m^2_i$'s
are the soft supersymmetry breaking Higgs boson masses and $\mu$ is the
supersymmetric Higgs mixing parameter; $\hat{m}^2_3=B\mu$; $Q_A$ are left--
handed squark doublets (the index $A$ numbers the generations), $U_A$ and $D_A$
are the up and down right--handed squark $SU(2)$ singlets, respectively; the
trilinear soft supersymmetry breaking scalar couplings are: $y^A_L=Y^A_lA_{L_A}
,~~y^A_D=Y^A_dA_{D_A}$ and $y^A_U=Y^A_uA_{U_A}$ where $Y$'s denote respective
Yukawa couplings; $g_i$'s are gauge couplings.

The scale $\Lambda\equiv M_{SUSY}$ has been chosen so that the term (A.8a)
vanishes (see the text).

1--loop corrections to the quartic Higgs couplings are as follows:
\newpage
$$\mbox{}\hspace*{-15em}4\lambda_1=g^2_1+g^2_2+\,\frac{1}{2(4\pi)^2}\sum_A
\left[\frac{8}{3}\,g^4_1\log\,\frac{m^2_{U_A}}{\Lambda^2}\right.$$
$$\mbox{}\hspace*{-13em}+\left({2\over{3}}g^4_1-8g^2_1Y^{A^2}_d+24\,Y^{A^4}_d\right)\,\log\,
\frac{m^2_{D_A}}{\Lambda^2}$$
$$\mbox{}\hspace*{-4.5em}\left.+\left(
3\,g^4_2+{1\over{3}}g^4_1+24\,Y^{A^4}_d-12\,g^2_2\,Y^{A^2}_d
-4\,g^2_1\,Y^{A^2}_d\right)\,\log\,\frac{m^2_{Q_A}}{\Lambda^2}\right]\eqno
\mbox{(A.13)} $$
\vspace*{10pt}\\
$$\mbox{}\hspace*{-15.5em}4\lambda_2=g^2_1+g^2_2+\,\frac{1}{2(4\pi)^2}\sum_A
\left[\frac{2}{3}\,g^4_1\log\,\frac{m^2_{D_A}}{\Lambda^2}\right.$$
$$\mbox{}\hspace*{-13.5em}+\left({8\over{3}}g^4_1-16g^2_1Y^{A^2}_u+24\,Y^{A^4}
_u\right)\,\log\,\frac{m^2_{U_A}}{\Lambda^2}$$
$$\mbox{}\hspace*{-5em}\left.+\left( 3\,g^4_2+{1\over{3}}g^4_1+24\,Y^{A^4}_u
-12\,g^2_2\,Y^{A^2}_u-4\,g^2_1\,Y^{A^2}_u\right)\,\log\,\frac{m^2_{Q_A}}
{\Lambda^2}\right]\eqno \mbox{(A.14)} $$
\vspace*{10pt}\\
$$\mbox{}\hspace{-4em}4\lambda_3=g^2_2-g^2_1\,+\,\frac{1}{(4\pi)^2}
\sum_A\left[
12\,Y^{A^2}_u Y^{A^2}_d\left( f\left( m^2_{D_A},m^2_{U_A}\right) \,+\,\log\,
\frac{m^2_{Q_A}}{\Lambda^2}\right) \right.$$
$$\mbox{}\hspace*{-1em}-\,g^2_1\left(\left( Y^{A^2}_u-Y^{A^2}_d\right)\log\,
\frac{m^2_{Q_A}}{\Lambda^2}\,-\,2\,Y^{A^2}_d\log\,\frac{m^2_{D_A}}{\Lambda^2}
\,-\,4\,Y^{A^2}_u\log\,\frac{m^2_{U_A}}{\Lambda^2}\right)$$
$$\mbox{}\hspace*{-10em}\left.-\,{1\over{6}}\,g^4_1\left(\log\,\frac
{m^2_{Q_A}}{\Lambda^2}\,+\,2\log\,\frac{m^2_{D_A}}{\Lambda^2}\,+\,8\log\,
\frac{m^2_{U_A}}{\Lambda^2}\right) \right.$$
$$\mbox{}\hspace*{-13em}\left.\,+\,\left(\frac{3}{2}g^4_2-3g^2_2\left(Y^{A^2}_u+Y^
{A^2}_d\right)\right)\log\,\frac{m^2_{Q_A}}{\Lambda^2}\right]\eqno
\mbox{(A.15)}
$$
\vspace*{10pt}\\
$$\mbox{}\hspace*{-7em}2\lambda_4=-\,g^2_2-\,\frac{1}{(4\pi)^2}\sum_A\left[
\left(\frac{3}{2}g^4_2-3\,g^2_2\left( Y^{A^2}_u+Y^{A^2}_d\right)\right)\log\,
\frac{m^2_{Q_A}}{\Lambda^2}\right. $$
$$\mbox{}\hspace*{-11.5em}\left.+\,\,6\,Y^{A^2}_uY^{A^2}_d\left(\log\,\frac
{m^2_{Q_A}}{\Lambda^2}\,+f\left( m^2_{D_A}, m^2_{U_A}\right)\right)\right]\eqno
\mbox{(A.16)} $$
\newpage
\newcommand{\be}{\begin{equation}}
\newcommand{\ee}{\end{equation}}
\begin{center}
{\sc {\bf APPENDIX B}}
\end{center}
\renewcommand{\theequation}{B.\arabic{equation}}
\setcounter{equation}{0}

\indent For easy reference we list here the 1--loop
supersymmetric renormalization group equations for the
Yukawa couplings and for the soft supersymmetry breaking
parame\-ters$^{[19]}$.
\be
\hspace*{-11em}2\,\frac{d}{dt}\,Y^A_l=Y^A_l\left(
3g^2_1+3g^2_2-3Y^{A^2}_l-\sum_B\left( Y^{B^2}_l+3Y^{B^2}_d\right)\right)
\ee
\vspace*{10pt}
\be
\hspace*{-2em}2\,\frac{d}{dt}\,Y^A_d=Y^A_d\left(
\frac{16}{3}\,g^2_3+\frac{7}{9}g^2_1+3g^2_2-3Y^{A^2}_d-\,Y^{A^2}_u-\sum_B\left(
Y^{B^2}_l
+3Y^{B^2}_d\right)\right)
\ee
\vspace*{10pt}
\be
\hspace*{-7em}2\,\frac{d}{dt}\,Y^A_u=Y^A_u\left(
\frac{16}{3}\,g^2_3+\,\frac{13}{9}\,g^2_1+3g^2_2-3Y^{A^2}_u-\,Y^{A^2}_d
-3\sum_B Y^{B^2}_u\right)
\ee
\vspace*{10pt}
\be
\hspace*{-10.5em}\frac{d}{dt}\,m^2_{L_A}=3g^2_2M^2_2+g^2_1M^2_1-Y^{A^2}_l\left(
m^2_{L_A}+
m^2_{E_A}+m^2_1+A^2_{L_A}\right)
\ee
\vspace*{10pt}
\be
\hspace*{-13.5em}\frac{d}{dt}\,m^2_{E_A}=4g^2_1M^2_1-2Y^{A^2}_l\left(
m^2_{L_A}+
m^2_{E_A}+m^2_1+A^2_{L_A}\right)
\ee
\vspace*{10pt}
\begin{eqnarray}
\hspace*{-3em}\frac{d}{dt}\,m^2_{Q_A}&=&\frac{16}{3}\,g^2_3M^2_3+3g^2_2M^2_2+\frac{1}{9}\,
g^2_1M^2_1-Y^{A^2}_d\left( m^2_{Q_A}+
m^2_{D_A}+m^2_1+A^2_{D_A}\right)\nonumber\\
\hspace*{-3em}&&-Y^{A^2}_u\left( m^2_{Q_A}+m^2_{U_A}+m^2_2+A^2_{U_A}\right)
\end{eqnarray}
\vspace*{10pt}
\be
\hspace*{-8em}\frac{d}{dt}\,m^2_{D_A}=\frac{16}{3}\,g^2_3M^2_3+\frac{4}{9}\,
g^2_1M^2_1-2Y^{A^2}_d\left( m^2_{Q_A}+m^2_{D_A}+m^2_1+A^2_{D_A}\right)
\ee
\vspace*{10pt}
\be
\hspace{-8em}\frac{d}{dt}\,m^2_{U_A}=\frac{16}{3}\,g^2_3M^2_3+\frac{16}{9}\,
g^2_1M^2_1-2Y^{A^2}_u\left( m^2_{Q_A}+m^2_{U_A}+m^2_2+A^2_{U_A}\right)
\ee
\vspace*{10pt}
\begin{eqnarray}
\hspace*{-8em}\frac{d}{dt}\,m^2_1&=&3\,g^2_2M^2_2+g^2_1M^2_1-3\sum_A
Y^{A^2}_d\left( m^2_{Q_A}+m^2_{D_A}+m^2_1+A^2_{D_A}\right)\nonumber\\
\hspace*{-8em}&&-\sum_AY^{A^2}_l\left(
m^2_{L_A}+m^2_{E_A}+m^2_1+A^2_{L_A}\right)
\end{eqnarray}
\vspace*{10pt}
\be
\hspace*{-9.5em}\frac{d}{dt}\,m^2_2=3\,g^2_2M^2_2+g^2_1M^2_1-3\sum_A
Y^{A^2}_u\left( m^2_{Q_A}+m^2_{U_A}+m^2_2+A^2_{U_A}\right)
\ee
\vspace*{10pt}
\be
\hspace*{-14.5em}\frac{d}{dt}\,\mu=\,\frac{1}{2}\,\mu\left(
3g^2_2+g^2_1-\sum_B\left(
Y^{B^2}_l+3Y^{B^2}_d+3Y^{B^2}_u\right)\right)
\ee
\vspace*{10pt}
\be
\hspace*{-7.5em}\frac{d}{dt}\,B=-\left( 3g^2_2M_2+g^2_1M_1\right)
-\sum_B\left( A_{L_B}Y^{B^2}_l+3A_{D_B}Y^{B^2}_d+3A^2_{U_B}Y^{B^2}_u\right)
\ee
\vspace*{10pt}
\be
\hspace*{-6.5em}\frac{d}{dt}\,A_{L_A}=-\left(3g^2_2M_2+g^2_1M_1\right)
-3A^2_{L_A}Y^{A^2}_l-\sum_B\left( A^2_{L_B}Y^{B^2}_l+3A_{D_B}Y^{B^2}_d\right)
\ee
\begin{eqnarray}
\hspace*{-6em}\frac{d}{dt}\,A_{D_A}&=&-\left(\frac{16}{3}g^2_3M_3+3g^2_2M_2
+\frac{7}{9}\,g^2_1M_1\right)-A_{U_A}Y^{A^2}_u-3A_{D_A}Y^{A^2}_d\nonumber\\
\hspace*{-6em}&&-\sum_B\left( A_{L_B}Y^{B^2}_l+3A_{D_B}Y^{B^2}_d\right)
\end{eqnarray}
\vspace*{10pt}
\begin{eqnarray}
\hspace*{-5em}\frac{d}{dt}\,A_{U_A}&=&-\left(\frac{16}{3}g^2_3M_3+3g^2_2
M_2+\frac{13}{9}\,g^2_1M_1\right)-3A_{U_A}Y^{A^2}_u-A_{D_A}Y^{A^2}_d\nonumber\\
\hspace*{-5em}&&-3\sum_BA_{U_B}Y^{B^2}_u
\end{eqnarray}
\vspace*{10pt}
\be
\hspace*{-29em}\frac{d}{dt}\left(\frac{M_i}{g^2_i}\right)=0
\ee
\vspace*{10pt}
\be
\hspace*{-30em}\frac{d}{dt}\,g_i=b_ig^3_i
\ee
\vspace*{25pt}\\
Here $t=\frac{1}{(4\pi)^2}\ln\frac{M^2_X}{Q^2}$ and $b_1=1+10n_G/3$,
{}~$b_2=5-2n_G$,~$b_3=-9+2n_G$.
The above equations are used in the range $(M_X,~~M_{SUSY})$. In the range
$(M_{SUSY},~~M_Z)$ (i.e. after decoupling of squarks and gluinos we use the
equations given in ref. [6].
\newpage
\begin{center}
{\sc {\bf REFERENCES}}
\end{center}
\vspace*{10pt}
\noindent \begin{itemize}
\item[{[1]}] 
L.E. Ib\'{a}nez and G.G. Ross, Phys. Lett. {\bf B110}, 215 (1982);\\
K. Inoue et al., Prog. Theor. Phys. {\bf 68}, 927 (1982);\\
L. Alvarez--Gaum\'{e}, J. Polchinsky and M. Wise, Nucl. Phys.
{\bf B221}, 495 (1983);\\
J. Ellis, J. Hagelin, D. Nanopoulos and K. Tamvakis, Phys. Lett.
{\bf B125}, 275 (1983);\\
L.E. Ib\'{a}nez and G. L\'{o}pez, Nucl. Phys. {\bf B233}, 511 (1984).
\item[{[2]}] 
R. Barbieri and G.F. Giudice, Nucl. Phys. {\bf B306}, 63 (1988);\\
G.G. Ross and R.G. Roberts, Nucl. Phys. {\bf B377}, 571 (1992).
\item[{[3]}] 
M. Dress and M.M. Nojiri, DESY preprint, Desy 92--101, Slac
Pub--5860, July 1992.
\item[{[4]}] 
G. Gamberini, G. Ridolfi and F. Zwirner, Nucl. Phys. {\bf B331},
331 (1990);\\
P.H. Chankowski, Phys. Rev. {\bf D41}, 2877 (1990).
\item[{[5]}] 
Y. Okada, M. Yamaguchi and T. Yanagida, Prog. Theor. Phys. {\bf
85}, 1 (1991);\\
H.E. Haber and R. Hempfling, Phys. Rev. Lett. {\bf 66}, 1815 (1991);\\
J. Ellis, G. Ridolfi and F. Zwirner, Phys. Lett. {\bf B257}, 83 (1991);\\
R. Barbieri, M. Frigeni and F. Caravaglio, Phys. Lett. {\bf
B258}, 395 (1991).
\item[{[6]}] 
P.H. Chankowski. Phys. Rev. {\bf D41}, 2877 (1990).
\item[{[7]}] 
M.A. Diaz and H.E. Haber, Phys. Rev. {\bf D45}, 4246 (1991);\\
K. Sasaki, M. Carena and C.E.M. Wagner, Nucl. Phys. {\bf B381},
66 (1992);\\
P.H. Chankowski, S. Pokorski and J. Rosiek, Phys. Lett. {\bf
B281}, 100 (1992);\\
H.E. Haber and R. Hempfling, Santa Cruz preprint SCIPP--91/33.
\item[{[8]}] 
J. Ellis, S. Kelly and D.V. Nanopoulos, Phys. Lett. {\bf 260B}, 131 (1991);\\
U. Amaldi, W. deBoer and M. Furstenau, Phys. Lett. {\bf 260B}, 447 (1991);\\
P. Langacker and M. Luo, Phys. Rev. {\bf D44}, 817 (1991);\\
F. Anselmo, L. Cifarelli, A. Peterman and A. Zichichi, Nuovo Cimento
{\bf 104A}, 1817 (1991);\\
G.G. Ross and R.G. Roberts, Nucl. Phys. {\bf B377}, 571 (1992).
\item[{[9]}] 
P. Langacker and N. Polonsky, University of Pennsylvania
preprint, UPR--0513T.
\item[{[10]}] 
M. Carena, S. Pokorski and C.E.M. Wagner, Max--Planck--Institute preprint,
MPI--Ph/93--10.
\item[{[11]}] 
H. Arason, D.J. Castano, B. Keszthelyi, S. Mikaelian, E.J.
Piard, P. Ramond and B.D. Wright, University of Florida preprint
UFIFT--HEP--91--33.
\item[{[12]}] 
J.R. Espinosa and M. Quir\'{o}s, Phys. Lett. {\bf B266}, 389 (1991).
\item[{[13]}] 
M. Carena, T.E. Clark, C.E.M. Wagner, W.A. Bardeen, K.
Sasaki, Nucl. Phys. {\bf B369}, 33 (1992).
\item[{[14]}] 
A. Bouquet, J. Kaplan and C.A. Savoy, Nucl. Phys. {\bf B262},
299 (1985).
\item[{[15]}] 
J.M. Fr\`{e}re, D.R.T. Jones and S. Raby, Nucl. Phys. {\bf
B222}, 11 (1983);\\
M. Claudson, L. Hall and I. Hinchliffe, Nucl. Phys. {\bf B228},
501 (1983);\\
J.P. Derendinger and C.A. Savoy, Nucl. Phys. {\bf B237}, 307 (1984).
\item[{[16]}] 
M. Olechowski and S. Pokorski, Phys. Lett. {\bf B214}, 393 (1988);\\
S. Pokorski, "How is the isotopic spin symmetry of quark masses
broken ?"  Proceedings of the XIIth International Workshop on
Weak Interactions and Neutrinos, Ginosar, Israel, April 1989 and
"On weak isospin breaking in the quark mass spectrum" in
Festschrift for Leon Van Hove, World Scientific, Singapore, 1990.
\item[{[17]}] 
See e.g. K. Griest, M. Kamionkowski and M. Turner, Phys. Rev.
{\bf D41}, 3565 (1990).
\item[{[18]}] 
M. Drees and K. Hikasa, Phys. Lett. {\bf B252}, 127 (1990).
\item[{[19]}] 
H.P. Nilles, "Beyond the Standard Model", in Proceedings of the 1990
Theoretical Advanced Study Institute in Elementary Particle Physics, p. 633;
Eds. M. Cvetic and P. Langacker, World Scientific;\\
P.H. Chankowski, Diploma Thesis (1990), University of Warsaw;\\
A. Seidl, Diploma Thesis (1990), Technical University, Munich;\\
W. Majerotto and B. M\"{o}sslacher, Z. Phys. {\bf C48}, 273 (1990);\\
S. Bertolini, F. Borzumati, A. Masiero and G. Ridolfi, Nucl. Phys.
{\bf B353}, 591 (1991);\\
M. Drees and M.M. Nojiri, Nucl. Phys. {\bf B369}, 54 (1992);\\
B. Ananthanarayan, G. Lazarides and Q. Shafi, Bartol Research Institute
preprint BA--92--29.
\item[{[20]}] 
K. Inoue, A. Kakuto, H. Komatsu and S. Takeshita, Prog. Theor. Phys. {\bf 67},
889 (1982) and {\bf 68}, 927 (1983);\\
B. Gato, J. Leon, J. Perez--Mercader and M. Quiros, Nucl. Phys. {\bf B253},
285 (1985);\\
N.K. Falck, Z. Phys. {\bf C30}, 247 (1986).
\item[{[21]}] 
See for instance, G,C. Branco, A.J. Buras and J.M. G\,{e}rard,
Nucl. Phys. {\bf B259}, 306 (1985);\\
P. Krawczyk and S. Pokorski, Phys. Rev. Lett. {\bf 60}, 182 (1988);\\
S. Pokorski, "How is the isotopic spin symmetry of quark masses
broken ?" Proceedings of the XIIth Intern. Workshop on Weak
Interactions and Neutrinos, Ginosar, Israel, April 1989.
\item[{[22]}] 
S. Dimopoulos, L.J. Hall and S. Raby, Phys. Rev. Lett. {\bf 68}.
1984 (1992); Phys. Rev. {\bf D45}, 4192 (1992);\\
G.W. Anderson, S. Raby, S. Dimopoulos and L. Hall, preprint
OHSTPY--HEP--92--018, October 1992.
\item[{[23]}] 
P. Gondolo, M. Olechowski and S. Pokorski, "Neutralinos as dark
matter in supergravity models", preprint MPI--Ph/92--81,
September 1992, to appear in Proceedings of the XXVI
International Conference on High Energy Physics, Dallas, August 1992;\\
Max--Planck--Institute for Physics preprint, to be published.
\item[{[24]}] 
R. Arnowitt and P. Nath, Phys. Rev. Lett. {\bf 69}, 725 (1992);\\
P. Nath and R. Arnowitt, Phys. Lett. {\bf B289}, 368 (1992).
\end{itemize}
\newpage
\begin{center}
{\large {\bf TABLE 1}}
\end{center}
\vspace*{20pt}
\begin{center}
Coefficients in Eqs. (2.15--2.17) for $M_{SUSY}=300~\mbox{GeV}$.\\
The numbers in brackets are without radiative corrections $(M_{SUSY}=M_Z)$.\\
\end{center}
\vspace*{50pt}
\begin{center}
\begin{tabular}{||r||ccccc||}\hline\hline
$\hat{m}_1^2$~~~~~~~~~~~~~~&$M^2_o$&$A^2_o$&$m^2_o$&$\mu^2_o$&$M_oA_o$\\
\hline\hline
$\tan\beta=2$&.478&$-$.002&.996&1.461&.005\\
{}~&(.478)&($-$001)&(.997)&(1.407)&(.004)\\
\underline{$M_t=130$}~~~~~$\tan\beta=10$&.286&$-$.041&.931&1.541&.092\\
{}~&(.222)&($-$.022)&(.928)&(1.495)&(.087)\\
$\tan\beta=30$&$-$.961&$-$.223&.438&1.220&.441\\
{}~&($-$1.401)&($-$.116)&(.414)&(1.184)&(.452)\\
\hline
$\tan\beta=2$&.480&$-$.002&.996&.444&.004\\
{}~&(.482)&($-$.001)&(.997)&(.411)&(.003)\\
\underline{$M_t=180$}~~~~~$\tan\beta=10$&.316&$-$.038&.939&.963&.079\\
{}~&(.267)&($-$.018)&(.937)&(.906)&(.070)\\
$\tan\beta=30$&$-$.765&$-$.215&.503&.761&.400\\
{}~&($-$1.128)&($-$.101)&(.484)&(.717)&(.381)\\
\hline\hline
$\hat{m}_2^2$~~~~~~~~~~~~~~&$M^2_o$&$A^2_o$&$m^2_o$&$\mu^2_o$&$M_oA_o$\\
\hline\hline
$\tan\beta=2$&$-$1.580&$-$.118&.296&1.406&.478\\
{}~&($-$2.176)&($-$.125)&(.260)&(1.407)&(.574)\\
\underline{$M_t=130$}~~~~~$\tan\beta=10$&$-$1.235&$-$.110&.434&1.499&.445\\
{}~&($-$1.756)&($-$.118)&(.403)&(1.495)&(.545)\\
$\tan\beta=30$&$-$1.198&$-$.102&.436&1.218&.413\\
{}~&($-$1.688)&($-$.108)&(.408)&(1.184)&(.499)\\
\hline
$\tan\beta=2$&$-$2.767&$-$.022&$-$.401&.409&.088\\
{}~&($-$3.523)&($-$.021)&($-$.435)&(.411)&(.095)\\
\underline{$M_t=180$}~~~~~$\tan\beta=10$&$-$2.406&$-$.085&$-$.129&.906&.343\\
{}~&($-$3.121)&($-$.084)&($-$.167)&(.906)&(.386)\\
$\tan\beta=30$&$-$2.340&$-$.072&$-$.123&.733&.292\\
{}~&($-$3.017)&($-$.070)&($-$.155)&(.717)&(.322)\\
\hline\hline
\end{tabular}
\end{center}
\begin{center}
\begin{tabular}{||r||ccccc||}\hline\hline
$m_Q^2$~~~~~~~~~~~~~~&$M^2_o$&$A^2_o$&$m^2_o$&$\mu^2_o$&$M_oA_o$\\
\hline\hline
$\tan\beta=2$&5.568&$-$.042&.767&.000&.175\\
{}~&(6.358)&($-$.042)&(.772)&(.000)&(.193)\\
\underline{$M_t=130$}~~~~~$\tan\beta=10$&5.623&$-$.044&.796&.000&.185\\
{}~&(6.413)&($-$.046)&(.780)&(.000)&(.210)\\
$\tan\beta=30$&5.196&$-$.066&.659&.000&.281\\
{}~&(5.898)&($-$.068)&(.636)&(.000)&(.313)\\
\hline
$\tan\beta=2$&5.142&$-$.008&.524&.000&.033\\
{}~&(5.910)&($-$.007)&(.521)&(.000)&(.033)\\
\underline{$M_t=180$}~~~~~$\tan\beta=10$&5.219&$-$.034&.605&.000&.145\\
{}~&(5.974)&($-$.033)&(.593)&(.000)&(.151)\\
$\tan\beta=30$&4.866&$-$.051&.489&.000&.216\\
{}~&(5.546)&($-$.050)&(.471)&(.000)&(.231)\\
\hline\hline
$m_U^2$~~~~~~~~~~~~~~&$M^2_o$&$A^2_o$&$m^2_o$&$\mu^2_o$&$M_oA_o$\\
\hline\hline
$\tan\beta=2$&4.395&$-$.083&.536&.000&.348\\
{}~&(5.019)&($-$.083)&(.507)&(.000)&(.383)\\
\underline{$M_t=130$}~~~~~$\tan\beta=10$&4.640&$-$.077&.629&.000&.323\\
{}~&(5.299)&($-$.079)&(.602)&(.000)&(.363)\\
$\tan\beta=30$&4.669&$-$.071&.631&.000&.299\\
{}~&(5.344)&($-$.072)&(.605)&(.000)&(.332)\\
\hline
$\tan\beta=2$&3.535&$-$.015&.048&.000&.065\\
{}~&(4.121)&($-$.014)&(.043)&(.000)&(.063)\\
\underline{$M_t=180$}~~~~~$\tan\beta=10$&3.803&$-$.060&.242&.000&.252\\
{}~&(4.389)&($-$.056)&(.222)&(.000)&(.258)\\
$\tan\beta=30$&3.853&$-$.051&.247&.000&.213\\
{}~&(4.458)&($-$.047)&(.230)&(.000)&(.215)\\
\hline\hline
\end{tabular}
\end{center}
\vspace*{30pt}
\begin{center}
\begin{tabular}{||r||ccc||}\hline\hline
$\hat{m}_3^2$~~~~~~~~~~~~~~&$\mu_oM_o$&$\mu_oA_o$&$\mu_oB_o$\\
\hline\hline
$\tan\beta=2$&$-$.070&$-$.273&1.186\\
{}~&(.004)&($-$.294)&(1.186)\\
\underline{$M_t=130$}~~~~~$\tan\beta=10$&$-$.143&$-$.252&1.222\\
{}~&($-$.073)&($-$.273)&(1.223)\\
$\tan\beta=30$&$-$.182&$-$.408&1.088\\
{}~&($-$.305)&($-$.427)&(1.088)\\
\hline
$\tan\beta=2$&.290&$-$.305&.641\\
{}~&(.346)&($-$.307)&(.641)\\
\underline{$M_t=180$}~~~~~$\tan\beta=10$&.265&$-$.378&.951\\
{}~&(.352)&($-$.390)&(.952)\\
$\tan\beta=30$&.444&$-$.463&.847\\
{}~&(.558)&($-$.471)&(.847)\\
\hline\hline
\end{tabular}
\end{center}
\newpage
\begin{center}
{\large {\bf TABLE 2}}
\end{center}
\vspace*{20pt}
\begin{center}
Generic spectra of the supersymmetric
particles with the lightest\\pseudoscalar $A$ or neutralino or chargino
(underlined).
\end{center}
\vspace*{45pt}
\begin{center}
\begin{tabular}{|l|rrrrrr|} \hline\hline
$M_t$&130&130&130&130&130&130\\
$\tan\beta$&2&2&2&10&10&10\\ \hline\hline
$M_A$&\underline{230.0}&420.0&380.0&\underline{190.0}&300.0&220.0\\
$m_Q$&225.0&250.0&250.0&300.0&375.0&425.0\\
$m_U$&185.0&150.0&170.0&260.0&315.0&385.0\\
\hline
$M_{\mbox{$h$}}$&58.6&60.0&60.4&94.2&95.5&96.0\\
$M_{\mbox{$n_1$}}$&39.8&\underline{32.5}&32.9&48.0&\underline{45.7}&48.0\\
$M_{\mbox{$n_2$}}$&87.4&73.3&74.9&85.4&75.9&80.9\\
$M_{\mbox{$n_4$}}$&191.9&283.9&248.9&207.0&176.9&177.2\\
$M_{\mbox{$c_1$}}$&49.5&47.0&\underline{45.1}&76.4&64.9&\underline{60.6}\\
$M_{\mbox{$c_2$}}$&222.2&304.1&271.9&217.2&187.0&183.7\\
$M_{\mbox{$\tilde{t}_1$}}$&145.6&84.6&123.6&177.7&242.3&399.9\\
$M_{\mbox{$\tilde{t}_2$}}$&302.6&325.0&322.3&389.2&454.0&440.0\\
$M_{\mbox{$\tilde{q}$}}$&228.6--245.7&253.7--313.5&253.8--302.9&305.5--322.2&
379.4--422.8&428.9--443.1\\
$M_{\mbox{$\tilde{l}$}}$&82.8--91.0&247.5--253.0&233.3--239.0&86.7--113.5
&283.9--301.2&186.4--234.3\\
\hline\hline
$M_t$&180&180&180&180&180&180\\
$\tan\beta$&2&2&2&30&30&30\\ \hline\hline
$M_A$&\underline{300.0}&550.0&355.0&\underline{175.0}&535.0&185.0\\
$m_Q$&225.0&300.0&225.0&300.0&525.0&325.0\\
$m_U$&165.0&160.0&145.0&260.0&345.0&285.0\\
\hline
$M_{\mbox{$h$}}$&75.6&79.9&74.5&111.7&117.7&113.0\\
$M_{\mbox{$n_1$}}$&37.7&\underline{22.6}&32.9&47.4&\underline{22.5}&52.1\\
$M_{\mbox{$n_2$}}$&83.5&47.6&74.7&86.6&43.8&95.1\\
$M_{\mbox{$n_4$}}$&231.0&379.6&251.0&244.3&407.6&250.6\\
$M_{\mbox{$c_1$}}$&51.6&63.7&\underline{45.3}&88.5&\underline{45.2}&97.1\\
$M_{\mbox{$c_2$}}$&256.1&373.5&273.7&246.5&411.2&252.4\\
$M_{\mbox{$\tilde{t}_1$}}$&233.1&138.1&224.4&232.1&361.6&258.9\\
$M_{\mbox{$\tilde{t}_2$}}$&285.5&392.5&281.6&398.7&563.9&418.2\\
$M_{\mbox{$\tilde{q}$}}$&229.2--260.4&303.4--386.9&229.4--275.0&
284.5--353.6&524.3--809.2&309.9--371.5\\
$M_{\mbox{$\tilde{l}$}}$&137.2--143.4&332.9--339.0&194.1--200.6&89.6--177.4&
549.3--798.4&\underline{82.5}--174.4\\
\hline\hline
\end{tabular}
\end{center}
\newpage
\begin{center}
{\large {\bf Table 3}}
\end{center}
\vspace*{20pt}
\begin{center}
Comparison of the supersymmetric particle spectra,\\
for the same sets of the input parameters,\\with and without (the numbers
in brackets) radiative corrections
\end{center}
\vspace*{45pt}
\begin{center}
\begin{tabular}{|l|rrrr|} \hline\hline
$M_t$&130&130&180&180\\
$\tan\beta$&2&10&2&30\\ \hline\hline
$M_A$&233.0&228.0&465.0&315.0\\
$m_Q$&225.0&425.0&400.0&325.0\\
$m_U$&185.0&385.0&320.0&225.0\\
\hline
$M_{\mbox{$h$}}$&58.6 (52.0)&96.0 (88.9)&90.1 (54.0)&111.1 (90.8)\\
$M_{\mbox{$n_1$}}$&38.5 (40.4)&56.2 (65.1)&60.3 (60.2)&32.1 (51.8)\\
$M_{\mbox{$n_2$}}$&85.4 (87.9)&89.5 (117.5)&115.5 (114.1)&61.3 (100.3)\\
$M_{\mbox{$n_4$}}$&184.6 (208.1)&186.1 (244.6)&396.0 (423.6)&292.0 (426.0)\\
$M_{\mbox{$c_1$}}$&46.4 (52.7)&74.8 (109.0)&134.0 (131.2)&59.3 (98.8)\\
$M_{\mbox{$c_2$}}$&215.7 (236.1)&193.7 (253.6)&385.7 (414.6)&298.3 (430.3)\\
$M_{\mbox{$\tilde{t}_1$}}$&149.4 (143.5)&344.8 (333.0)&262.7 (241.5)&228.4
(159.1)\\
$M_{\mbox{$\tilde{t}_2$}}$&300.8 (304.8)&484.3 (495.0)&500.4 (512.4)&399.1
(433.3)\\
$M_{\mbox{$\tilde{q}$}}$&228.8--246.7&428.8--441.6&402.5--439.0
&323.6--494.0\\
{}~&(228.4--245.1)&(428.9--439.5)&(402.0--438.9)&(312.0--534.4)\\
$M_{\mbox{$\tilde{l}$}}$&103.9--111.6&197.9--214.8&159.2--185.3&324.2--452.2\\
{}~&(41.9--55.7)&(72.5--121.3)&(70.7--120.7)&(305.2--415.1)\\
$\max|\Delta|$&12.7 (23.1)&8.2 (12.3)&23.7 (84.3)&29.9 (75.9)\\
\hline\hline
\end{tabular}
\end{center}
\newpage
\begin{center}
{\sc {\bf FIGURE CAPTIONS}}
\end{center}
\vspace*{20pt}
\begin{itemize}
\item[Fig. 1.]
The regions in the unification scale parameters ($m_o,~~M_o$),
obtained by demanding radiative gauge symmetry breaking,
in form of the scatter plots obtained by the scanning procedure
described in text. These are 2--dimensional projections
of the solutions in the 5--dimensional parameter space.
Crosses: solutions already excluded by experimental constraints
discussed in the text; squares: solutions compatible with experimental
constraints with $\max |\Delta_{ij}|<30$; dots: solutions
compatible with experimental constaints with $30<\max |\Delta_{ij}|\\
<100$. Dashed contours: solutions with $\max\ |\Delta_{ij}|<10$.
Solid contours: solutions compatible with experimental constraints
and with $\max |\Delta_{ij}|<30$, but with no radiative corrections.\\
A) $M_t=130$ GeV, $\tan\beta=2$; B) $M_t=130$ GeV, $\tan\beta=10$;\\
C) $M_t=180$ GeV, $\tan\beta=2$; D) $M_t=180$ GeV, $\tan\beta=30$.\\
In case C) there are no solutions corresponding to the contour criteria.
\item[Fig. 2.]
Same as Fig. 1  for ($m_o,~~A_o$).
\item[Fig. 3.]
Same as Fig. 1 for ($A_o~~M_o$).
\item[Fig. 4.]
Same as Fig. 1 for ($B_o/m_o,~~A_o/m_o$).
\item[Fig. 5.]
Same as Fig. 1 for the {\it low energy} parameters ($M_2,~~\mu$).
\item[Fig. 6.]
Same as Fig. 1 for the physical lightest neutralino  mass $M_n$
and the pseudoscalar mass $M_A$.
\item[Fig. 7.]
Same as Fig. 1 for the gaugino content of the lightest
neutralino,\\ $Z_n=Z^2(1,1)+Z^2(1,2)$ and its mass $M_n$.
\item[Fig. 8.]
Same as Fig. 1 for the lighter chargino mass $M_c$ and the
pseudoscalar mass $M_A$.
\item[Fig. 9.]
Same as Fig. 1 for the lighter chargino mass $M_c$ and the
lightest neutralino mass $M_n$.
\item[Fig. 10.]
Same as Fig. 1 for the physical lightest stop mass
$m_{\tilde{t}}$ and the universal gaugino mass at $M_X,~~M_o$.
\item[Fig. 11.]
Same as Fig. 1 for the physical stau mass $m_{\tilde{\tau}}$ and
the universal scalar mass at $M_X,~~m_o$.
\item[Fig. 12.]
Same as Fig. 1 for the stop mass $m_{\tilde{t}}$ and the
pseudoscalar mass $M_A$.
\item[Fig. 13.]
Same as Fig. 1 for the chargino mass $M_c$ and the stop mass
$m_{\tilde{t}}$.
\item[Fig. 14.]
Solutions for $M_t=130~\mbox{GeV}$ and $\tan\beta=30$. Notation as in
Fig. 1 (dots correspond now to solutions with $30<\max |\Delta_{ij}|$
$<250$). The corresponding value of $Y_t/Y_b$ at $M_X$ is 1.08.
\item[Fig. 15.]
Solutions for $M_t=130~\mbox{GeV}$ and $\tan\beta=31$. There only
exist solutions with $\max|\Delta_{ij}|>30$. Plotted are solutions
with $\max|\Delta_{ij}|<500$. The corresponding values of $Y_t/Y_b$ at
$M_X$ is 1.04.
\item[Fig. 16.]
Same as Fig. 15 for $M_t=180~\mbox{GeV}$ and $\tan\beta=50$. The
corresponding values of $Y_t/Y_b$ at $M_X$ is 1.18.
\item[Fig. 17.]
Same as Fig. 15 for $M_t=180~\mbox{GeV}$ and $\tan\beta=51$. The
corresponding values of $Y_t/Y_b$ at $M_X$ is 1.12.
Plotted are solutions with $\max|\Delta_{ij}|<750$.
\end{itemize}
\end{document}